\documentstyle[12pt, axodraw]{article}
\newcommand{\bc} {\begin{center}}
\newcommand{\ec} {\end{center}}
\newcommand{\be} {\begin{equation}}
\newcommand{\ee} {\end{equation}}
\newcommand{\Section}[1]{\section{#1} \setcounter{equation}{0}}
\newcommand{\Sectionn}[1]{\section*{#1} \setcounter{equation}{0}}
\def\lsim{\mathrel{\rlap{\lower4pt\hbox{\hskip1pt$\sim$}}
    \raise1pt\hbox{$<$}}}         

\def\gsim{\mathrel{\rlap{\lower4pt\hbox{\hskip1pt$\sim$}}
    \raise1pt\hbox{$>$}}}         

\newcommand{\overstar}[1]{\mathrel{\rlap{\hbox{$#1$}}
    \raise3pt\hbox{$\hspace{0.5mm}^*$}}}
\makeatletter
\def\appendix{\par\clearpage
  \setcounter{section}{0}
  \setcounter{subsection}{0}
  \@addtoreset{equation}{section}
  \def\@sectname{Appendix~}
  \def\theequation{\thesection.\arabic{equation}}
  \def\thesection{\Alph{section}}}
\makeatother

\renewcommand{\theequation}{\thesection.\arabic{equation}}

\begin{document}

\begin{titlepage}

\vskip 3.5cm \centerline{\bf  NON-FORWARD NLO BFKL
KERNEL$^{~\ast}$}
\vskip 1cm \centerline{  V.S. Fadin$^{\dagger}$ and  R. Fiore$^{\ddagger}$ }

\vskip 1cm

\centerline{\sl $^{\dagger}$  Budker Institute for
Nuclear Physics and Novosibirsk State University,}
\centerline{\sl  630090 Novosibirsk, Russia}  \vskip .3cm
\centerline{\sl $^{\ddagger}$ Dipartimento di Fisica,
Universit\`a della Calabria} \centerline{\sl and Istituto
Nazionale di Fisica Nucleare, Gruppo collegato di
Cosenza,} \centerline{\sl I-87036 Arcavacata di Rende,
Cosenza, Italy}

\vskip 1cm

\begin{abstract}
Details of the calculation of the non-forward BFKL kernel at
next-to-leading order (NLO) are offered. Specifically we show
the calculation of the two-gluon production contribution.
This contribution was the last missing part of the kernel.
Together with the NLO gluon Regge trajectory, the NLO contribution
of one-gluon production and the contribution of
quark-antiquark production which were found before it
defines the kernel completely for any colour state in the $t$-channel,
in particular the Pomeron kernel presented recently.

\end{abstract}
\vskip 1cm \vfill \hrule \vskip.3cm \noindent
$^{\ast}${\it Work supported in part by the Ministero Italiano
dell'Istruzione, dell'Universit\`a e della Ricerca, in
part by INTAS and
in part by the Russian Fund of Basic Researches.}\\
\vskip 0.5cm \vfill $\begin{array}{ll}
^{\dagger}\mbox{{\it email address:}} & \mbox{fadin@inp.nsk.su} \\
^{\ddagger}\mbox{{\it email address:}} & \mbox{fiore@cs.infn.it} \\
\end{array}$
\end{titlepage}
\eject

\section{Introduction}

Talking about the  BFKL kernel one usually has in mind
the kernel of the BFKL equation~\cite{BFKL} for the case
of forward scattering, i.e. for the momentum transfer
$t=0 $ and  vacuum quantum numbers in the $t$-channel.
However, the BFKL approach is not limited to this
particular case and, what is more, from the beginning it
was developed for arbitrary $t$ and for all possible
$t$-channel colour states. Initially it was done in the
leading logarithmic approximation (LLA), which means
summation  of terms of the type $[\alpha _{s}\ln s]^{n}$
($\alpha _{s}$ is the QCD coupling constant and $s$ is
the squared c.m.s. energy). This approximation can
provide only qualitative results, since it does not fix
scales, neither for energy nor for transverse momenta
determining the running coupling constant $\alpha _{s}$.
Therefore calculation of radiative corrections to LLA
seems to be a daily need. Unfortunately, till now it is
not completed, although the forward BFKL kernel in the
NLO  was found already five years ago~\cite{FL98}.

The problem of the development of the BFKL approach in the
next-to leading approximation (NLA) is naturally  divided
into two parts, in compliance with the representation of
scattering amplitudes in this approach by the convolution
of the impact factors of interacting particles with the
Green's function of two Reggeized gluons in the
$t$-channel. The impact factors describing  the scattering of
particles by the Reggeized gluons contain all the
dependence on the nature of the particles and are energy
independent.  All the dependence on energy is defined by the
universal (i.e. process independent) Green's function, which
is determined by the BFKL kernel. For a consistent
description of scattering amplitudes one needs to
know the impact factors with the same accuracy as the
kernel. Especially interesting is the highly virtual
photon impact factor, because it can be calculated
from the ``first principles" in perturbative QCD. Unfortunately,
this calculation turned out to be a very complicated
problem, which is not yet solved, although a noticeable
progress has been reached here~\cite{photon impact factor}.
Recently an important step was done finding the solution of a
related problem: the NLO impact factor for the transition of a
virtual photon in a light vector meson was calculated in
the case of $t=0$ and longitudinal
polarizations~\cite{Ivanov:2004pp}.  The NLO impact
factors are known also  at parton level (i.e. for quarks
and gluons)~\cite{parton impact factors}.

The calculation of the NLO BFKL kernel for the
non-forward scattering was not  completed until recently.  We
remind that for any colour group
representation ${\cal R}$ in the $t$-channel the kernel
is given by the the sum of
``virtual" and ``real" parts~\cite{FF98}. The
``virtual" part is universal (i.e. it does not
depend on ${\cal R}$) and is expressed through the NLO
gluon Regge trajectory \cite{trajectory}.  The ``real"
part is related to the particle production in
Reggeon-Reggeon collisions and consists of
one-gluon, two-gluon and quark-antiquark contributions.
The first contribution is expressed through the
effective Reggeon-Reggeon-gluon NLO vertex \cite{vertex}.
Apart from a colour  coefficient, it also is universal and known
\cite{gluon octet kernel}.  Each of last two contributions
is written as a sum of two terms  with
depending on $R$ coefficients, at that only one of these
terms enters in the kernel for the antisymmetric
colour octet representation ${\cal
R}=8_a$ (gluon channel), whereas the kernel for the colour
singlet representation ${\cal R}=1$ (Pomeron channel) contains
both terms.
For the case of quark-antiquark production both these
terms are known \cite{quark part of the kernel}.
Instead, only the piece related to the  gluon channel was known
for the case of two-gluon production \cite{gluon
octet kernel}. Note that for scattering of physical
(colourless) particles only the Pomeron channel exists.
Nevertheless the gluon channel plays an important role.
It is caused by the possibility to use this channel for a
check of self-consistency, and, finally, for a proof of
the gluon Reggeization (see Ref.~\cite{proof of
Reggeization} and references therein).

Thus, the two-gluon production contribution
was the only missing piece in the the non-forward BFKL
kernel. Now it is calculated and the Pomeron kernel is
known \cite{FF-2004}. Here we present the details
of the calculation of the two-gluon contributions and the
non-forward BFKL kernel at NLO for all possible colour states
in the $t$-channel. Since the quark
contribution to the non-forward kernel is known
~\cite{quark part of the kernel} for any ${\cal R}$, we
shall consider in the following only the gluon
contribution, i.e. we shall work in pure gluodynamics.

In the next Section we present the gluon piece of the
gluon trajectory, the general form of the ``real"
contribution to the kernel and its part related  to
one-gluon production. In Section 3 we derive the
contribution to the kernel from the two-gluon production
and define  the ``symmetric'' part of this contribution.
The colour group relations used in this Section are given
in Appendix A. The ``symmetric'' part of  the two-gluon
contribution  is considered in Section 4.  The three
pieces contributing to this part are calculated in
Appendices B, C and D, respectively. Finally, in Section
5 the non-forward kernel is discussed.

\Section{The ``virtual'' and ``one-gluon'' parts of the kernel}

As usual, we  utilize the Sudakov decomposition of momenta,
denoting $p_1$ and $p_2$  the light-cone vectors close to the initial
particle momenta $p_A$ and $p_B$ respectively, so that
$2p_1p_2=(p_A+p_B)^2=s$.
We use the conventional dimensional regularization with
the space-time dimension $D=4+2\epsilon$ and the
normalization adopted in Ref.~\cite{FF98}. The BFKL
equation for the Mellin transform
$G_{\omega}$ of the Green's function $G$ is written as
\begin{equation}\label{Green function}
\omega G^{\left( {\cal R}\right) }_{\omega}\left( \vec
q_1,\vec q_2;\vec q\,\right) = \vec q_1^{\,2}\vec
q_1^{\,\prime\,2}\delta ^{\left( D-2\right) }\left( \vec
q_1-\vec q_2\right) + \int\frac{d^{D-2}r}{\vec
r^{\:2}(\vec r-\vec q)^2}{\cal K}^{\left( {\cal R}\right)
}\left( \vec q_1,\vec r;\vec q\right)G^{\left( {\cal
R}\right) }_{\omega}\left( \vec r,\vec q_2;\vec
q\,\right) ~.
\end{equation}
Here  $q_i$ and $q_i^\prime \equiv q_i-q\,, \; (i=1\div
2)$ are the Reggeon (Reggeized gluon)  momenta, $q\simeq
q_\perp$ is the momentum transfer; $q^2\simeq
q^2_\perp =-\vec q^{~2}=t;\;$ the vector sign is used
for denoting components of momenta transverse to the
$p_1p_2$ plane. The BFKL kernel
${\cal K}^{\left( {\cal R}\right) }$ has the form
\begin{equation}
{\cal K}^{\left( {\cal R}\right) }\left( \vec q_1,\vec
q_2;\vec q\,\right) = \left[ \omega \left( -\vec
q_1^{\,2}\right) +\omega \left( - \vec q_1^{\,\prime\,2}
\right)\right] \vec q_1^{\,2}\vec q_1^{\,\prime\,2}\delta
^{\left( D-2\right) }\left( \vec q_1-\vec q_2\right) +
{\cal K}_r^{\left( {\cal R}\right) }\left( \vec q_1,\vec
q_2;\vec q\right) ~,   \label{kernel=virtual+real}
\end{equation}
i.e. it is given by the sum of the ``virtual" part, determined by
the gluon Regge trajectory $\,\omega(t)$ (actually the
trajectory is $j(t)=1+\omega(t))$,  and the ``real" part,
related to particle production in Reggeon-Reggeon
collisions. In the limit $\epsilon\rightarrow
0$~\cite{trajectory} the trajectory is given by
\begin{equation}
\omega(t) =\omega^{(1)}(t)
\left\{1+\frac{\omega^{(1)}(t)}
{4}\left[\frac{11}{3}+\left(2\zeta(2)-
\frac{67}{9}\right) \epsilon+\left(\frac{404}{27}
-2\zeta(3)\right) \epsilon^2 \right]\right\}~, \label{NLO
trajectory}
\end{equation}
where $\omega^{(1)}(t)$ is the one-loop contribution,
whose expression is
\begin{equation}
\omega^{(1)}(t) =\frac{g^2N_c t}{2(2\pi)^{D-1}}
\int\frac{d^{D-2}r}{\vec r^{\,2}\vec
r^{\,\prime\,2}}=-g^2 \frac{N_c \Gamma(1-\epsilon)}{(4
\pi)^{D/2}} \frac{\Gamma^2(\epsilon)}{\Gamma(2\epsilon)}
(\vec q^{\,2})^\epsilon \,. \label{LO trajectory}
\end{equation}
Here and in the following $\vec r^{\,\prime}=\vec r-\vec
q~,\;$$N_c$ is the number of colors, $\Gamma(x)$ is the
Euler function, $\zeta(n)$ is the Riemann zeta function,
($\zeta(2)=\pi^2/6$)  and $g$ is the bare coupling
constant, concerned with the renormalized coupling $g_\mu
$ in the ${\overline{MS}}$ scheme through the relation
\begin{equation}
g=g_\mu \mu ^{-\mbox{\normalsize $\epsilon$}}\left[ 1+
\frac{11}3 \frac{\bar g_\mu ^2}{2\epsilon }\right]
~;\;\;\;\;\bar g_\mu ^2=\frac{g_\mu ^2N\Gamma (1-\epsilon
)}{(4\pi )^{2+{\epsilon }}}~. \label{coupling
renormalization}
\end{equation}
The ``real" part ${\cal K}_r^{\left( {\cal R} \right) }$
of the kernel is related to real particle production in
Reggeon-Reggeon collisions. It can be presented in the
form \cite{FF98}
\[
{\cal K}^{\left( {\cal R}\right) }_{r}(\vec{q}_{1},\vec
q_{2}; \vec q)={\cal K}^{{\left( {\cal R}\right)
}\Lambda}_{r} (\vec{q}_{1},\vec q_{2}; \vec q)
 -\frac{1}{2}\int
\frac{d^{D-2}r}{\vec{r}^{~2} \vec r^{\,\prime\,2}}{\cal
K}^{{\left( {\cal R}\right) }B}_{r}(\vec{q}_{1},\vec r;
\vec q)
\]
\begin{equation}
\times {\cal K}_{r}^{{\left( {\cal R}\right)
}B}(\vec{r},\vec{q}_{2}; \vec q) \, \ln \left(
\frac{s^2_{\Lambda }}
{(\vec{q}_1-\vec{r})^2(\vec{q}_2-\vec{r})^2}\right)~,\label{real
kernel=unsubtracted-subtraction}
\end{equation}
where  the ``non-subtracted"  kernel ${\cal
K}_r^{{\left( {\cal R}\right) }\Lambda}$ is
\begin{equation}
{\cal K}_r^{{\left( {\cal R}\right) }\Lambda}
(\vec{q}_{1},\vec q_{2}; \vec q)= \frac{\langle
bb^{\prime }|\hat {{\cal P}}_{ {\cal R}}|aa^{\prime
}\rangle }{n_{{\cal R}}}\sum_{ J } \int \gamma_{ab}^{J
}\left( q_{1},q_{2}\right) \left( \gamma_{a^{\prime
}b^{\prime }}^{ J} \left( q_{1}^{\prime},q_{2}^{\prime
}\right) \right)^{\ast } \frac{d\phi_{J}}
{2(2\pi)^{D-1}}~.\label{unsubtracted kernel}
\end{equation}
Here $\hat {{\cal P}}_{ {\cal R}}$ is the operator for
projection of two-gluon colour states on the
representation $ {\cal R}$; $a,a^{\prime}$ and
$b,b^{\prime}$ are Reggeon colour indices; $n_{\cal R}$
is the number of independent states in $ {\cal R}$;
$\gamma_{ab}^{J }\left( q_{1},q_{2}\right)$ is the
effective vertex for production of the state $J$ in the
collision of Reggeons with momenta $q_1=\beta p_1
+q_{1\perp}, \;\; q_2=-\alpha p_2 +q_{2\perp}$;
$d\phi_{J}$ is the corresponding phase space element;
the sum is over all possible states $J$. For a state $J$
consisting of particles with momenta $k_i$, with the
total momentum $k=q_1-q_2$, we have
\begin{equation}
d\phi_{J}=\frac{d k^2}{2\pi}\theta(s_\Lambda
-k^2)(2\pi)^{D} \delta^D(k-\sum_i
k_i)\prod_i\frac{d^{D-1}k_{i}}{\left( 2\pi
\right)^{D-1}2\epsilon _{i}}~. \label{phase space
element}
\end{equation}
The intermediate parameter $s_{\Lambda }$ in
Eq.~(\ref{real kernel=unsubtracted-subtraction}) must be
taken tending to infinity. The second term in the R.H.S.
of Eq.~(\ref{real kernel=unsubtracted-subtraction})
appears only at the NLO and serves for subtraction of the
large $k^2$ contribution, in order to avoid a double
counting of this region. At the leading order (LO) only
one-gluon production does contribute, so that $k^2=0$,
Eq.~(\ref{unsubtracted kernel}) does not depended on
$s_\Lambda$ and  gives the kernel in the leading (Born)
order:
\[
{\cal K}^{(\cal R)B}_r\left( \vec q_1,\vec q_2;\vec
q\,\right) =\frac{\langle bb^{\prime }|\hat {{\cal P}}_{
{\cal R}}|aa^{\prime }\rangle }{2n_{{\cal
R}}(2\pi)^{D-1}}\sum_{ G } \gamma_{ab}^{G }\left(
q_{1},q_{2}\right) \left( \gamma_{a^{\prime }b^{\prime
}}^{ G} \left( q_{1}^{\prime},q_{2}^{\prime }\right)
\right)^{\ast }
\]
\begin{equation}
= \frac{g^2N_c  c_{\cal R}}{(2\pi )^{D-1}}\left(
\frac{\vec q_1^{\,2}\vec q_2^{\,\prime\, 2}+\vec
q_2^{\,2}\vec q_1^{\,\prime\,2}}{(\vec q_1- \vec
q_2)^2}-\vec q^{\,2}\right)~.  \label{real born kernel}
\end{equation}
Here
\begin{equation}
c_{\cal R}=\frac{Tr\left(\hat {{\cal P}}_{{\cal
R}}T^d\bigotimes T^{d \ast}\right)}{N_c n_R}
  =\frac{\langle bb^{\prime }|\hat {{\cal P}}_{{\cal
R}}|aa^{\prime }\rangle }{N_c n_{{\cal
R}}}T^d_{ab}T^d_{b^{\prime }a^{\prime
}}\label{coefficients cR}
\end{equation}
are the group coefficients,  $T^d$ are the colour group
generators in the adjoint representation,
$T^d_{ab}=-if_{dab}$,  $f_{dab}$ are the group structure
constants. The projection operators and the coefficients
$c_{\cal R}$ for all possible representations ${\cal R}$
are given, for completeness, in the Appendix A. As it was
already noted, the most interesting representations are
the colour singlet (Pomeron channel, ${\cal R}=1$ ) and
the antisymmetric colour octet (gluon channel ${\cal
R}=8_a$). Respectively, we have for them
\begin{equation}
 c_1=1\,,\,\,\,~~~c_{8_a}=\frac{1}{2}\,. \label{born group koefficients}
\end{equation}
Note that for the symmetric colour octet ${\cal R}={8_s}$
we have $c_{8_s}=c_{8_a}$, so that at  LO the BFKL
kernels for the symmetric and anti-symmetric  octet
representations coincide. Since the trajectory
$\omega(t)$ is the eigenvalue of ${\cal K}^{(8_a)}$
(``bootstrap" of the gluon Reggeization), the same is
true for ${\cal K}^{(8_s)}$ (trajectory degeneration).
However, generally speaking, it does not mean that
amplitudes with ${\cal R}=8_s$ in the $t$-channel have
the Reggeized form (with the same trajectory but positive
signature), because such a form requires impact factors
proportional to the eigenfunction corresponding to the
eigenvalue $\omega(t)$. It turns out that for parton
scattering amplitudes in  LLA it is just the case.

Having the  one-gluon production vertices at
NLO~\cite{vertex}, one can easily calculate  the
one-gluon contribution to the kernel with the NLO
accuracy. Retaining only terms giving non-vanishing
contributions in the $\epsilon\rightarrow 0$ limit after
integration of the kernel over $d^{D-2}k$ in a
neighbourhood of the singular point $\vec k = \vec
q_1-\vec q_2=0$, we have~\cite{gluon octet kernel}
\[
{\cal K}_{G}^{({\cal R})}(\vec q_1,\vec q_2; \vec{q})
        =  \frac{g^2N_c c_{\cal R}}{(2\pi)^{D-1}}
\left\{ \left( \frac{\vec{q}_1^{\,2}
\vec{q}_2^{\,\prime\,2}+ \vec{q}_1^{\,\prime\,2}
\vec{q}_2^{\,2}}{\vec k^{\,2}}-
\vec{q}^{\,2}\right)\right.
\]
\[
\times \left(\frac{1}{2}+ \frac{g^2N_c
\Gamma(1-\epsilon)}{2(4\pi)^{2+\epsilon}} \left[ - (\vec
k^{\,2})^\epsilon \,
\left(\frac{2}{\epsilon^2}-\pi^2+4\,\epsilon\,\zeta(3)\right)
-\ln^2\left(\frac{\vec{q}_1^{\,2}}{
\vec{q}_2^{\,2}}\right)\right] \right)
\]
\[
+\frac{g^2N_c
\Gamma(1-\epsilon)}{6(4\pi)^{2+\epsilon}}
\left(\left[\frac{ \vec{q}_1^{\,\prime\,2}-
\vec{q}_2^{\,\prime\,2}}{\vec{q}_1^{\,2}-
\vec{q}_2^{\,2}}-\frac{ \vec{k}^{\,2}}{(\vec{q}_1^{\,2}-
\vec{q}_2^{\,2})^2} \left(\vec{q}_1^{\,2}+
\vec{q}_2^{\,2}+4
\vec{q}_1^{\,\prime}\vec{q}_2^{\,\prime}-2
\vec{q}^{\,2}\right)\right]
\right.
\]
\[
\times\left[\frac{2\vec{q}_1^{\,2}
\vec{q}_2^{\,2}}{{\vec{q}_1^{\,2}-
\vec{q}_2^{\,2}}}\;\ln\left(\frac{\vec{q}_1^{\,2}}{
\vec{q}_2^{\,2}}\right)-\vec{q}_1^{\,2}-
\vec{q}_2^{\,2}\right] +11\left[\frac{2\vec{q}_1^{\,2}
\vec{q}_2^{\,2}}{{\vec{q}_1^{\,2}- \vec{q}_2^{\,2}}}
+\frac{\vec{q}_1^{\,2} \vec{q}_2^{\,\prime\,2}-
\vec{q}_1^{\,\prime\,2} \vec{q}_2^{\,2}}{ \vec{k}^{\,2}}
- \frac{\vec{q}_1^{\,2}+
\vec{q}_2^{\,2}}{\vec{q}_1^{\,2}- \vec{q}_2^{\,2}}\,
\vec{q}^{\,2}\right]\, \ln\left(\frac{\vec{q}_1^{\,2}}{
\vec{q}_2^{\,2}} \right)
\]
\begin{equation}
-
2\vec{q}_1^{\,\prime}\vec{q}_2^{\,\prime}\Biggr)\Biggr\}
+ \Biggl ( \vec q_i
\leftrightarrow \vec q_i^{\,\prime} \Biggr )\,.
\label{one gluon contribution}
\end{equation}
For arbitrary $D$ this part of the kernel can be found in
the last of Refs.~\cite{vertex} (see there Eq.~(4.10)).
Note that the exchange $\vec q_i \leftrightarrow \vec
q_i^{\,\prime}$ implies also $\vec q \leftrightarrow
-\vec q$.  Since the colour structure of the one-gluon
production vertex is not changed at NLO, the coefficients
$c_{\cal R}$ here are the same as in Eq.~(\ref{real born
kernel}).

The remarkable properties of the kernel,  subsequent from
general arguments, are
\begin{equation}
{\cal K}^{({\cal R})}_{r}( 0,\vec{q}_{2};\vec{q}\,) =
{\cal K}^{({\cal R})}_{r}( \vec{q}_{1},0;\vec{q}\,)
={\cal K}^{({\cal R})}_{r}( \vec{q},\vec{q}_{2};\vec{q}
\,) ={\cal K}^{({\cal R})}_{r}(
\vec{q}_{1},\vec{q};\vec{q}\,) =0
 \label{gauge invariance of the kernel}
\end{equation}
and
\begin{equation}
{\cal K}^{({\cal R})}_{r}(
\vec{q}_{1},\vec{q}_{2};\vec{q} \,)= {\cal K}^{({\cal
R})}_{r}( -\vec{q} _{1}^{\, \prime},-\vec{q}_{2}^{\,
\prime };\vec{q}\,) ={\cal K}^{({\cal R})}_{r}(
-\vec{q}_{2},-\vec{q}_{1};- \vec{q}\,) ~. \label{symmetry
of the kernel}
\end{equation}
Properties (\ref{gauge invariance of the kernel}) mean
that the kernel turns into zero at zero transverse
momenta of the Reggeons and appear as a consequence of
the gauge invariance. Properties (\ref{symmetry of the
kernel}) are a consequence of the  crossing invariance
and the gluon identity.

The symmetry properties (\ref{symmetry of the kernel})
of ${\cal K}_{G}^{({\cal R})}(\vec q_1,\vec q_2; \vec{q})$
are evident from Eq.~(\ref{one gluon contribution}). Instead
properties  (\ref{gauge invariance of the kernel}) are not
so evident, but can be easily checked.

\Section{The two-gluon  production contribution}

The new states which appear in the sum over $J$ in
Eq.~(\ref{unsubtracted kernel}) at NLO are the  two-gluon
ones. For these (and only for these) states the integral
in Eq.~(\ref{unsubtracted kernel}) is logarithmically
divergent  at large $k^2$. In this region the two-gluon
production vertex factorizes into the product of the
one-gluon vertices (see below), so that the dependence on
$s_{{\Lambda }}$ in Eq.~(\ref{real
kernel=unsubtracted-subtraction}) is
cancelled~\cite{FF98}.

Lorentz-  and gauge- invariant representation of the
two-gluon production vertex has been obtained in
Ref.~\cite{LF89}. It has the form
\begin{equation}
\gamma_{i
j}^{G_1G_2}(q_1,q_2)=g^2\overstar{e}_{1}^{\,\mu}
\overstar{e}_{2}^{\,\nu}\left[\left(T^{d_1}
T^{d_2}\right)_{i j} A_{\mu\nu} (k_1,k_2) +\left(T^{d_2}
T^{d_1}\right)_{i j} A_{\nu\mu} (k_2,k_1)\right]~,
\label{gamma_GG covariant}
\end{equation}
where $e_i$ and $d_i$ are  gluon polarization vectors and
colour indices, respectively. Note that the tensor $A_{\mu\nu}
(k_1,k_2)$ depends not only on $k_1,k_2$, as it is
explicitly indicated, but on $p_1, q_1$ and $p_2, q_2$ as
well. The explicit expression of the tensor is given in
Ref.~\cite{LF89}. Its important property is the
Abelian-type gauge invariance
\begin{equation}
k_1^\mu A_{\mu\nu}(k_1,k_2)=k_2^\nu
A_{\mu\nu}(k_1,k_2)=0~.
\end{equation}
Another important  property of the tensor
$A_{\mu\nu}(k_1,k_2)$ is its transformation law under
simultaneous exchange $(p_1, q_1, q,  i)
\longleftrightarrow (p_2, -q_2, -q,  j)\,$, which we will
call $L\longleftrightarrow R$ exchange:
\begin{equation}
A_{\mu\nu}(k_1,k_2)|_{L\longleftrightarrow R}=
A_{\nu\mu}(k_2,k_1)~. \label{A under LR exchange}
\end{equation}
This property and the representation (\ref{gamma_GG
covariant}) guarantee that the vertex $\gamma_{i
j}^{G_1G_2}(q_1,q_2)$ is invariant with respect to the
$L\longleftrightarrow R$ exchange, as it must be. Taking
into account the representation (\ref{unsubtracted
kernel}), this invariance provides the symmetry of the
non-subtracted kernel with respect to the exchange $\vec
q_1\leftrightarrow -\vec q_2~,\;\;\vec q\leftrightarrow
-\vec q~$.

We use the Sudakov decomposition for the produced gluon
momenta $k_1$ and $k_2\;$ ($k=k_1+k_2=q_1-q_2$) in the
form ($i=1,2$)
\begin{equation}
k_i=\beta_i p_1 +\alpha_i p_2 +k_{i\perp}~,\;\;
s\alpha_i\beta_i =-k_{i\perp}^2=\vec k_i^{~2}~,\;\;
\beta_i=x_i \beta~, \;\; x_1+x_2=1~.
\label{decomposition of k_i}
\end{equation}
We find that it is convenient to use the same light-cone
gauge $e_ip_2= e_ik_i=0$ for both gluons, so that we put
\begin{equation}
e_=e_{i\perp} -\frac{(e_{i\perp}
k_{i\perp})}{(k_ip_2)}p_2~. \label{axial gauge}
\end{equation}
In this  gauge the vertex takes the form
\begin{equation}
\gamma_{i
j}^{G_1G_2}(q_1,q_2)=4g^2\overstar{e}_{1\perp}^{\alpha}
\overstar{e}_{2\perp}^{\beta} \left[\left(T^{d_1}
T^{d_2}\right)_{i j} b_{\alpha\beta} (q_1;k_1,k_2)
+\left(T^{d_2} T^{d_1}\right)_{i j} b_{\beta\alpha}
(q_1;k_2,k_1)\right]~, \label{gamma_GG}
\end{equation}
with the tensor
\begin{equation}
4b^{\alpha\beta}(q_1;k_1,k_2)=(g_{\perp}^{\alpha\mu}-\frac{k^{\alpha}
_{1\perp}p^{\mu}_2}{(p_2k_1)})(g_{\perp}^{\beta\nu}-\frac{k^{\beta}
_{2\perp}p^{\nu}_2}{(p_2k_2)})A_{\mu\nu}(k_1,k_2)~,\;\;\label{b
from A}
\end{equation}
The explicit form of this tensor in terms of the
Sudakov variables has been found in Ref.~\cite{bootstrap in
QMRK}:
\[
b^{\alpha\beta}(q_1;k_1,k_2)=
\frac{1}{2}g^{\alpha\beta}_{\perp}\left[
\frac{1}{k^2}\left(2q_{1\perp}\Lambda_{\perp}+q^2_{1\perp}
\frac{\Lambda_\perp
(2x_1x_2k_\perp-\Lambda_\perp(x_1-x_2))}{\Sigma}\right)
\right.
\]
\[
\left.-x_2\frac{q^2_{1\perp}-2q_{1\perp}k_{1\perp}}{\tilde
t_1}\right]-\frac{x_2k^{\alpha}_{1\perp}q^{\beta}_{1\perp}
-x_1q^{\alpha}_{1\perp}(q_1-k_1)^{\beta}_{\perp}}{x_1\tilde
t_1}
-\frac{q^2_{1\perp}k^{\alpha}_{1\perp}(q_1-k_1)^{\beta}
}{k^2_{1\perp}\tilde t_1}
\]
\begin{equation}
-\frac{x_1q^{\alpha}_{1\perp}\Lambda^{\beta}_{\perp}
+x_2\Lambda^{\alpha}_{\perp}q^{\beta}_{1\perp}}{x_1x_2
k^2}
-\frac{x_1q^2_{1\perp}k^{\alpha}_{1\perp}k^{\beta}_{2\perp}}
{k^2_{1\perp}\Sigma}-\frac{q^{2}_{1\perp}}{k^2\Sigma}(\Lambda^{\alpha}_{\perp}
k^{\beta}_{2\perp}+k^{\alpha}_{1\perp}\Lambda^{\beta}_{\perp})~,
\label{b}
\end{equation}
where $g_{\perp}^{\mu\nu}$ is the metric tensor in the
transverse plane:
\begin{equation}\label{transverse metric tensor}
g_{\perp}^{\mu\nu} = g^{\mu\nu} -
\frac{p_1^{\mu}p_2^{\nu} + p_2^{\mu}p_1^
{\nu}}{(p_1p_2)}~.
\end{equation}
Moreover, the following positions hold:
\[
\Lambda_\perp=(x_2k_1-x_1k_2)_\perp~,\;\;k^2=-\frac{\Lambda^2_\perp}{x_1x_2}~,\;\;
\Sigma=-(x_1k^2_{2\perp}+
x_2k^2_{1\perp})=-\Lambda_\perp^2-x_1x_2k^2_\perp~,\;\;
\]
\begin{equation}
\tilde
t_1=(q_1-k_1)^2=\frac{1}{x_1}(x_1(q_1-k_1)^2_{\perp}+
x_2k^2_{1\perp})~,\;\; \tilde t_2=(q_1-k_2)^2=
\frac{1}{x_2}(x_2(q_1-k_2)^2_{\perp}+ x_1k^2_{2\perp})~.
\label{denotations}
\end{equation}
Note that on account of Eq.~(\ref{b from A}) the tensors
$b^{\alpha\beta}(q_1;k_1,k_2)$ and
$b^{\beta\alpha}(q^{\prime}_1;k_2,k_1)$ contain in the
denominators ``extra" powers of $x_1$ and $x_2$, in
comparison with ``naive" expectations based on the
consideration of Feynman diagrams.

We use the notations which coincide with those adopted in
Ref.~\cite{gluon octet kernel}. Note, however, that in
Ref.~\cite{gluon octet kernel} two different gauges were
used for the two gluons: the gauge (\ref{axial gauge})
for the second gluon (that with momentum $k_2$), and the
gauge similar to the gauge (\ref{axial gauge}) with $p_2$ replaced
by $p_1$ for the first one (that with momentum $k_1$).
Therefore our tensors are related to those used in
Ref.~\cite{gluon octet kernel} by the equalities
\begin{equation}
b^{\alpha\beta}(q_1;k_1,k_2)=
\Omega^{\alpha\gamma}(k_{1}) c_{\gamma}^{\;\;
\beta}(k_1,k_2)~,\;\; b^{\beta\alpha}(q_1;k_2,k_1)=
\Omega^{\beta\gamma}(k_{2}) c_{\;\gamma}^{\prime\;
\;\alpha}(k_2,k_1)~,
\end{equation}
where the tensors
\begin{equation}
\Omega^{\alpha\beta}(k)= g_{\perp}^{\alpha\beta} -
2\frac{k_{\perp}^{\alpha}k_{\perp}^{\beta}}{k_{\perp}^2}~,\;\;\label{gauge
transformation Omega}
\end{equation}
with the property
\begin{equation}
\Omega^{\alpha\gamma}(k)\Omega_\gamma^{\;\;\beta}(k)=
g_{\perp}^{\alpha\beta}~, \label{Omega-squared}
\end{equation}
realize the transformations between the gauges with  the
gauge fixing vectors $p_1$ and $p_2$.

Let us consider the behaviour of the vertex (\ref{gamma_GG})
in the multi-Regge kinematics, i.e. in the limits
$x_1\rightarrow 1~, x_2\rightarrow 0$ and $x_1\rightarrow
0~, x_2\rightarrow 1$. The first of them corresponds to
the case  when the first gluon is much closer to the
particle $A$ in rapidity space than the second gluon.
Therefore in this limit the two-gluon production vertex
must be factorized as
\begin{equation}
\gamma^{G_1G_2}_{ij}(q_1,q_2)=\gamma^{G_1}_{il}(q_1,q_1-k_1)
\frac{1}{(q_1-k_1)^2_{\perp}}\gamma^{G_2}_{lj}(q_1-k_1,q_2)
~, \label{factorization 2=1X1}
\end{equation}
where $\gamma^{G}_{ij}(q_1,q_2)$ is the one-gluon production
vertex.  Indeed, at $x_1=1~, x_2=0 $ we have
\begin{equation}
\Sigma=-k^2_{2\perp}~, \;\; x_2 k^2=-k^2_{2\perp}~,
\;\;\tilde t_1=(q_1-k_1)^2_{\perp}~,\;\; x_2\tilde
t_2=k^2_{2\perp}~, \;\;
\end{equation}
that gives
\begin{equation}
b^{\beta\alpha}(q_1;k_2,k_1)|_{x_1=1} =0~,
\end{equation}
so that
\begin{equation}
\gamma_{ij}^{G_1G_2}(q_1,q_2)=4g^2\overstar{e}_{1\perp}^{\,\alpha}
\overstar{e}_{2\perp}^{\,\beta}\left(T^{d_1}
T^{d_2}\right)_{i j} b_{\alpha\beta}
(q_1;k_1,k_2)|_{x_1=1}~, \label{regge limit 1}
\end{equation}
where
\begin{equation}
b^{\alpha\beta}(q_1;k_1,k_2)|_{x_1=1}=\frac{1}{(q_1-k_1)^2_{\perp}}
\left[q_{1\perp}-\frac{q^2_{1\perp}}{k^2_{1\perp}}k_{1\perp}
\right]^{\alpha}\left[q_{1\perp}-k_{1\perp}-\frac{(q_{1\perp}
-k_{1\perp})^2}{k^2_{2\perp}}k_{2\perp} \right]^{\beta}~.
\label{b(k1k2) at x=1}
\end{equation}
Since in the gauge (\ref{axial gauge}) we have
\begin{equation}
\gamma^{G_1}_{ij}(q_1,q_1-k_1)=-2g\overstar{e}_{1\perp}^{\,\alpha}
T^{d_1}_{i
j}\left[q_{1\perp}-\frac{q^2_{1\perp}}{k^2_{1\perp}}k_{1\perp}
\right]_{\alpha}~,
\end{equation}
we see that the factorization property
(\ref{factorization 2=1X1}) is fulfilled. Moreover, with
account of the result (\ref{b(k1k2) at x=1}), from this
property it follows that
\[
\left(\frac{2g^2N_c  c_{\cal R}}{(2\pi
)^{D-1}}\right)^2\left(b^{\alpha\beta}(q_1;k_1,k_2)
b_{\alpha\beta}(q^{\prime}_1;k_1,k_2)\right)
|_{x_1=1}=
\]
\begin{equation}
\frac{{\cal K}^{({\cal R})B}_r\left( \vec q_1,\vec
q_1-\vec k_1;\vec q\,\right){\cal K}^{({\cal
R})B}_r\left(\vec q_1-\vec k_1, \vec q_2;\vec
q\,\right)}{(\vec q_1-\vec k_1)^2(\vec q^{~\prime}_1-\vec
k_1)^2}~. \label{bb|x=1=KK}
\end{equation}
This equality can be obtained also directly from
Eq.~(\ref{b(k1k2) at x=1}) using the expression
(\ref{real born kernel}) for the LO kernel.

In the second limit, i.e. $x_1=0~, x_2=1 $ we get
\[
\Sigma=-k^2_{1\perp}~, \;\; x_1 k^2=-k^2_{1\perp}~, \;\;
x_1\tilde t_1=k^2_{1\perp}~,\;\; \tilde
t_2=(q_1-k_2)^2_{\perp}~,
\;\;b^{\alpha\beta}(q_1;k_1,k_2)=0~,
\]
\begin{equation}
b^{\beta\alpha}(q_1;k_2,k_1)|_{x_2=1}=\frac{1}{(q_1-k_2)^2_{\perp}}
\left[q_{1\perp}-\frac{q^2_{1\perp}}{k^2_{2\perp}}k_{2\perp}
\right]^{\beta}\left[q_{1\perp}-k_{2\perp}-\frac{(q_{1\perp}
-k_{2\perp})^2}{k^2_{1\perp}}k_{1\perp}
\right]^{\alpha}~ \label{b(k1k2) at x=0}
\end{equation}
and
\begin{equation}
\gamma^{G_1G_2}_{ij}(q_1,q_2)=\gamma^{G_2}_{il}(q_1,q_1-k_2)
\frac{1}{(q_1-k_2)^2_{\perp}}\gamma^{G_1}_{lj}(q_1-k_2,q_2)
~.\label{factorization at x=0}
\end{equation}
We have also
\[
\left(\frac{g^2N_c  c_{\cal R}}{2(2\pi
)^{D-1}}\right)^2\left(b^{\alpha\beta}(q_1;k_2,k_1)b_{\alpha\beta}(q^{\prime}_1;k_2,
k_1)\right) |_{x_2=1}
\]
\begin{equation}
=\frac{{\cal K}^{({\cal R})B}_r\left( \vec q_1,\vec
q_1-\vec k_2;\vec q\,\right){\cal K}^{({\cal
R})B}_r\left(\vec q^{~\prime}_1-\vec k_2, \vec q_2;\vec
q\,\right)}{(\vec q_1-\vec k_2)^2(\vec q^{~\prime}_1-\vec
k_2)^2}~. \label{bb|x=0=KK}
\end{equation}

With the help of Eqs.~(\ref{unsubtracted kernel}) and
(\ref{gamma_GG})  the two-gluon contribution to the
``non-subtracted" kernel is presented in the form
\begin{equation}
{\Large {\cal K}_{GG}^{{\left( {\cal R}\right)
}\Lambda}\left( \vec q_1,\vec q_2;\vec q\,\right)
}=8g^4N_c^2\int \left(a_R F_a(k_1,k_2)+b_R
F_b(k_1,k_2)\right)\frac{d\phi_{GG}}{(2\pi)^{D-1}}~,
\label{K in terms of F}
\end{equation}
where the group coefficients $a_R$ and $b_R$ are defined
as
\begin{equation}
a_R=\frac{Tr\left(\hat {{\cal P}}_{{\cal
R}}(T^{d_1}T^{d_2})\bigotimes
(T^{d_1}T^{d_2})^{\ast}\right)}{N^2_c n_R}~,\;\; b_R=
\frac{Tr\left(\hat {{\cal P}}_{{\cal
R}}(T^{d_1}T^{d_2})\bigotimes
(T^{d_2}T^{d_1})^{\ast}\right)}{N^2_c n_R}~,
\label{coefficients a and b}
\end{equation}
and the functions $F_a$ and $F_b$ as
\begin{equation}
F_a(k_1,k_2)=b^{\alpha\beta}(q_1;k_1,k_2)b_{\alpha\beta}
(q^{\prime}_1;k_1,k_2)+b^{\beta\alpha}(q_1;k_2,k_1)b_{\beta\alpha}(q^{\prime}_1;k_2,k_1)~,
\label{F_a}
\end{equation}
\begin{equation}
F_b(k_1,k_2)=b^{\alpha\beta}(q_1;k_1,k_2)b_{\beta
\alpha}(q^{\prime}_1;k_2,k_1)+
b^{\beta\alpha}(q_1;k_2,k_1)b_{\alpha\beta}(q^{\prime}_1;k_1,k_2)~.
\label{F_b}
\end{equation}
It is easy to see  that $a_R=c_R^2$, since
\[
{Tr\left(\hat {{\cal P}}_{{\cal
R}}(T^{d_1}T^{d_2})\bigotimes
(T^{d_1}T^{d_2})^{\ast}\right)}={Tr\left(\hat {{\cal
P}}_{{\cal R}}(T^{d_1}\bigotimes T^{{d_1\ast} })
(T^{d_2}\bigotimes T^{{d_2\ast} })\right)}=
\]
\begin{equation}
={Tr\left(\hat {{\cal P}}_{{\cal R} }T^{d_1}\bigotimes
T^{{d_1\ast}}\hat {{\cal P}}_{{\cal R} }T^{d_2}\bigotimes
T^{d_2{\ast}}\right)}=\frac{1}{n_R}{Tr\left(\hat {{\cal
P}}_{{\cal R} }T^{d_1}\bigotimes T^{{d_1\ast}}\right)}
{\left(\hat {{\cal P}}_{{\cal R} }T^{d_2}\bigotimes T^{
d_2{\ast}}\right)}~.
\end{equation}
This relation is important for the cancellation of the
$s_{\Lambda}$--dependence in the kernel (\ref{real
kernel=unsubtracted-subtraction}). Due to the result
(\ref{born group koefficients}) it gives, in particular,
\begin{equation}
a_0=1~,\;\;a_{8_a}=a_{8_s}=\frac{1}{4}~.\label{two-gluon
group koefficients}
\end{equation}
For the coefficients $b_R$, using the relation
\begin{equation}
(T^{d_1}T^{d_2})\bigotimes
(T^{d_1}T^{d_2})^{\ast}-(T^{d_1}T^{d_2})\bigotimes
(T^{d_2}T^{d_1})^{\ast}=if^{dd_2d_1}(T^{d_1}T^{d_2})\bigotimes
T^{d{\ast}}=\frac{N_c}{2} T^{d}\bigotimes T^{d{\ast}}~,
\end{equation}
with account of Eq.~(\ref{coefficients cR}) we obtain
\begin{equation}
b_R=a_R-\frac{1}{2}c_R=c_R\left(c_R-\frac{1}{2}\right)~,
\end{equation}
i.e. $b_1=1/2$, whereas for both symmetric and
antisymmetric colour octet representations the
coefficients $b_R$ are zero.  This is especially
important for the antisymmetric case, since the vanishing
of $b_{8_a}$ is crucial for the gluon Reggeization. Note
that the vanishing of $b_{8_s}$ means that in pure
gluodynamics the kernels for both octet representations
coincide at the NLO as well as at the LO.

In terms of the variables $k_{i\perp}$ and $x_i~$  the
phase space element $d\phi_{GG}$ can be written as
\begin{equation}
d\phi_{GG}=\frac{dx_1dx_2}{4x_1x_2}\delta(1-x_1-x_2)
\frac{d^{D-2}k_1d^{D-2}k_2}{(2\pi
)^{(D-1)}}\delta^{D-2}(k_{\perp}-k_{1\perp}-
k_{2\perp})\theta(s_\Lambda -k^2)~.\label{phase space in
x}
\end{equation}
Here the identity of the final gluons is taken into
account by the factor $1/2!$, so that integration must be
performed over all the phase space;  $k_{\perp}=
q_{1\perp}- q_{2\perp}= q_{1\perp}^{\,\prime}
-q_{2\perp}^{\,\prime}$ and  $k^2$ must be expressed in
terms of $x_i$ and $k_{i\perp}$ (see
Eq.~(\ref{denotations})). We recall that the parameter
$s_{_{\Lambda }}$ must be taken tending to infinity
before the limit $\epsilon\rightarrow 0$. From the
expression (\ref{b}) one can see that the tensors
$b^{\alpha\beta}$ fall down as $1/\vec k_i^{~2}$ at $\vec
k_i^{~2}\rightarrow \infty$ at fixed $x_i$.  Therefore
the integral over $\vec k_i$ in Eq.~(\ref{K in terms of
F}) is well  convergent  in the ultraviolet region, so
that the restrictions imposed by the theta-function can
be written as
\begin{equation}
 x_i \geq \frac{\vec k_{i}^{\:2}}{s_{_{\Lambda }}}~.
 \label{restrictions on x_i}
\end{equation}

Let us discuss the properties (\ref{gauge invariance of
the kernel}) and (\ref{symmetry of the kernel}) of the
two-gluon contribution to the kernel
(\ref{kernel=virtual+real}). As for the  subtraction
term, its properties (\ref{gauge invariance of the
kernel}), related to gauge invariance,  follow directly
from the corresponding properties of the ``Born" kernel
(\ref{real born kernel}).  Properties (\ref{symmetry of
the kernel}) are provided by the appropriate symmetries
expressed by Eq.~(\ref{real born kernel}) and the
invariance of the logarithm and the integration measure
$d^{D-2}r/(\vec r^{\,2}\vec r^{\,\prime\,2})$ in
Eq.~(\ref{real kernel=unsubtracted-subtraction}) under
the exchanges $\vec q_i\leftrightarrow -\vec
q_i^{\,\prime}~,\;\;\vec r\leftrightarrow -\vec
r^{\,\prime}$, as well as $\vec q_1\leftrightarrow -\vec
q_2~,\;\;\vec q\leftrightarrow -\vec q~,\;\;\vec
r\leftrightarrow -\vec r~$.

Turn now to the non-subtracted contribution~(\ref{K in
terms of F}).   Since its ``gauge-invariance" properties
(\ref{gauge invariance of the kernel}) are provided (see
Eqs.~(\ref{F_a}) and (\ref{F_b})) by the conversion into
zero of the tensor $b^{\alpha_1\alpha _2}(q_1;k_1,k_2)$
at $q_{1\perp}=0$ and at
$q_{2\perp}=(q_{1}-k_1-k_2)_\perp=0$, they can be easily
seen from the representation~(\ref{b}). The symmetry with
respect to the exchanges $\vec q_i\leftrightarrow -\vec
q_i^{\,\prime}$ can be also easily seen from the
representation ~(\ref{K in terms of F}), taking into
account Eqs.~(\ref{F_a}) and (\ref{F_b}).  As it was
already discussed,  the symmetry relative to the
exchanges $\vec q_1\leftrightarrow -\vec q_2~,\;\;\vec
q\leftrightarrow -\vec q~$ follows from the general
representation (\ref{unsubtracted kernel}) and the
invariance of the vertex $\gamma_{i j}^{G_1G_2}(q_1,q_2)$
with regard  to the $L\longleftrightarrow R$ exchange.
However,  it is not easy to derive this symmetry from the
representation ~(\ref{K in terms of F}) using
Eqs.~(\ref{F_a}), (\ref{F_b}) and the tensor
$b_{\alpha\beta}(q_1; k_1,k_2)$ presented in
Eq.~(\ref{b}). The matter is that the gauge (\ref{axial
gauge}) breaks the symmetry between $p_2$ and $p_1$, so
that a transformation law of this tensor  under the
$L\longleftrightarrow R$ exchange has not a simple form
like that of Eq.~(\ref{A under LR exchange}). Moreover,
the choice of the variables $x_i$ also destroys the
symmetry between $p_1$ and $p_2$. To restore the symmetry
one has to introduce the variables $y_i$, defined as
\begin{equation}
y_i=\frac{\alpha_i}{\alpha}~,\;\;\;\;  y_1+y_2=1~,  \label{y}
\end{equation}
concerned with $x_i$ by the relation
\begin{equation}
\frac{y_1}{y_2}=\frac{x_2}{x_1}\frac{k^2_{1\perp}}{k^2_{2\perp}}~.
\end{equation}
Using this relation and taking into account that
\begin{equation}
k^2=\frac{(x_2\vec k_1 -x_1\vec
k_2)^2}{x_1x_2}=\frac{(y_2\vec k_1 -y_1\vec
k_2)^2}{y_1y_2}~,
\end{equation}
it is easy to see that the phase space element
$d\phi_{GG}$ (\ref{phase space in x}) is invariant under the
exchange $x_i\leftrightarrow y_i$. Note that
in terms of the Sudakov variables the
$L\longleftrightarrow R$ exchange means
\begin{equation}
\beta_i\leftrightarrow \alpha_i~,\;\;\;
q_{1\perp}\leftrightarrow
-q_{2\perp}~,\;\;\;q_{\perp}\leftrightarrow -q_{\perp}
\end{equation}
or, in terms of $x_i$ and $y_i$,
\begin{equation}
x_i\leftrightarrow y_i~,\;\;\; q_{1\perp}\leftrightarrow
-q_{2\perp}~,\;\;\; q_{\perp}\leftrightarrow -q_{\perp}~.
\end{equation}
As it was already mentioned, under this exchange the
transformation law for the tensor $b^{\alpha\beta}$ is
not the same as for  $A^{\mu\nu}$ of Eq.~(\ref{A under LR
exchange}), because of the choice of the gauge
(\ref{axial gauge}). It is not difficult to understand
that the transformation low must be
\begin{equation}
b_{\alpha\beta}(q_1; k_1,k_2)|_{L\longleftrightarrow R}=
\Omega^{\alpha}_{\;\;\gamma}(k_1)\Omega^{\beta}_{\;\;\delta}(k_2)b_{\delta\gamma
}(q_1; k_2,k_1)~, \label{b under LR exchange}
\end{equation}
where the tensors $\Omega^{\alpha\beta}$ (see the
definition (\ref{gauge transformation Omega})) take into
account the gauge change under the $L\leftrightarrow R$
exchange. One can check  directly using  Eq.~(\ref{b})
that the property (\ref{b under LR exchange}) is indeed
fulfilled. Together with that of
Eq.~(\ref{Omega-squared}), this property demonstrates
once more that the functions $F_a$ and $F_b$, defined by
Eqs.~(\ref{F_a}) and (\ref{F_b}) respectively, are
invariant with respect to the $L\leftrightarrow R$
exchange. This means that, due to the invariance of the
phase space element discussed above, the contribution
(\ref{K in terms of F}) is symmetric under the
replacement  $\vec q_1 \leftrightarrow -\vec q_2, \;\;
\vec q \leftrightarrow -\vec q $.

From Eqs.~(\ref{real kernel=unsubtracted-subtraction}),
(\ref{bb|x=1=KK}), (\ref{bb|x=0=KK}) and (\ref{F_a}) it
follows that the subtraction  term can be written as
\[
\frac{g^4N_c^2
c_R^2}{(2\pi)^{D-1}}\int\frac{d^{2+2\epsilon}k_1}
{(2\pi)^{D-1}}
\left(F_a(k_1,k_2)|_{x_1=1}+F_a(k_1,k_2)|_{x_2=1}\right)
\ln\left(\frac{s_{\Lambda}^2}{k_1^2k_2^2}\right)
\]
\[
=8g^4N_c^2 a_R\int\frac{d\phi_{GG}}{(2\pi)^{D-1}}
\left(x_1F_a(k_1,k_2)|_{x_1=1}+x_2F_a(k_1,k_2)|_{x_2=1}\right)
\]
\begin{equation}
+\frac{g^4N_c^2a_R}{(2\pi)^{D-1}}\int\frac{d^{2+2\epsilon}k_1}
{(2\pi)^{D-1}}
\left(F_a(k_1,k_2)|_{x_1=1}-F_a(k_1,k_2)|_{x_2=1}\right)
\ln\left(\frac{k_2^2}{k_1^2}\right)~. \label{substraction
in terms of F}
\end{equation}
Here the equality $a_R=c^2_R$, the expression (\ref{phase
space in x}) for the phase space element and the
restrictions given by the inequality (\ref{restrictions
on x_i}) on $x_i$ were taken into account. Note that the
second integral in the R.H.S of Eq.~(\ref{substraction in
terms of F}) is completely antisymmetric with respect to
the substitution
 $\vec q_1 \leftrightarrow -\vec q_2, \;\; \vec q
\leftrightarrow -\vec q $. Therefore the subtraction term
can be obtained by symmetrization of the first integral.
Consequently, using the definition
\begin{equation}\label{integral +}
\left(\frac{f(x)}{x(1-x)}\right)_+\equiv
\frac{1}{x}[f(x)-f(0)] +\frac{1}{(1-x)}[f(x)-f(1)]~,
\end{equation}
we can write the two-gluon contribution to the  kernel
(\ref{real kernel=unsubtracted-subtraction}) in the limit
$s_{\Lambda}\rightarrow \infty$ in the form
\begin{equation}
{\cal K}^{({R})}_{GG}(\vec q_1,\vec q_2;\vec q)=
\frac{2g^4N_c^2}{(2\pi)^{D-1}} \hat{{\cal
S}}\int_{0}^{1}{dx}\int\frac{d^{2+2\epsilon}k_1}
{(2\pi)^{D-1}} \left(\frac{a_R F_a(k_1,k_2)+b_R
F_b(k_1,k_2)}{x(1-x)}\right)_+~, \label{kernel as +
integral}
\end{equation}
where $x\equiv x_1$ and the operator $\hat{{\cal S}}$
symmetrizes with respect to the
substitution $\vec q_1 \leftrightarrow -\vec q_2, \;\;
\vec q \leftrightarrow -\vec q $. It was used here that
$F_b(k_1,k_2)|_{x_1=0}=F_b(k_1,k_2)|_{x_1=1}=0$,
according to the definition (\ref{F_b}) and the
properties $b^{\alpha\beta}(q_1; k_1,
k_2)|_{x_1=0}=b^{\beta\alpha}(q_1; k_2,
k_1)|_{x_1=1}=0~$.

Since the coefficient $b_{R}$ is equal to zero for an
octet representation, the first term  in Eq.~(\ref{kernel
as + integral}) is determined by the two-gluon
contribution to the octet kernel, which is already
calculated~\cite{gluon octet kernel}. Therefore, our task
is to calculate the second term. However, we have found
that to calculate just this term is not the most
convenient way because of the rather complicated form of
the convolution $F_b$. Instead of this we find more
convenient to calculate the ``symmetric" contribution
\begin{equation}
{\cal K}^{({s})}_{GG}(\vec q_1,\vec q_2;\vec q)=
\frac{2g^4N_c^2}{(2\pi)^{D-1}} \hat{{\cal
S}}\int_{0}^{1}{dx}\int\frac{d^{2+2\epsilon}k_1}
{(2\pi)^{D-1}}
\left(\frac{F_s(k_1,k_2)}{x(1-x)}\right)_+~,
\label{symmetric kernel}
\end{equation}
where the function
\begin{equation}
F_s(k_1, k_2)=F_a(k_1,k_2)+ F_b(k_1,k_2)
\end{equation}
is given by the convolution
\begin{equation}
(b^{\alpha\beta}(q_1;k_1,k_2)+b^{\beta\alpha}(q_1;k_2,k_1))
(b_{\alpha\beta}(q^{\prime}_1;k_1,k_2)
+b_{\beta\alpha}(q^{\prime}_1;k_2,k_1))~.
 ~\label{Ae-Ad}
\end{equation}
Taking into account that $a_8=1/4$, we can present the
two-gluon contribution to the kernel for any
representation $R$ as
\begin{equation}
{\cal K}^{({R})}_{GG}(\vec q_1,\vec q_2;\vec
q)=4(a_R-b_R){\cal K}^{({8})}_{GG}(\vec q_1,\vec q_2;\vec
q) +b_R {\cal K}^{({s})}_{GG}(\vec q_1,\vec q_2;\vec q) .
\label{kernel through symmetric kernel}
\end{equation}

\Section{The ``symmetric" contribution to the kernel}
Calculation of ${\cal K}^{({s})}_{GG}(\vec q_1,\vec
q_2;\vec q)$ seems to be  more convenient since the sum
\[
b^{\alpha\beta}(q_1;k_1,k_2)+b^{\beta\alpha}(q_1;k_2,k_1)=
\frac{q^2_{1\perp}k^{\alpha}_{1\perp}k^{\beta}_{2\perp}}
{k^2_{1\perp}k^2_{2\perp}}
\]
\[-\frac{1}{2}g^{\alpha\beta}_{\perp}x_2\frac{q^2_{1\perp}-2q_{1\perp}k_{1\perp}}{\tilde
t_1}-\frac{x_2k^{\alpha}_{1\perp}q^{\beta}_{1\perp}
-x_1q^{\alpha}_{1\perp}(q_1-k_1)^{\beta}_{\perp}}{x_1\tilde
t_1}
-\frac{q^2_{1\perp}k^{\alpha}_{1\perp}(q_1-k_1)^{\beta}_{\perp}
}{k^2_{1\perp}\tilde t_1}
\]
\begin{equation}
-\frac{1}{2}g^{\alpha\beta}_{\perp}x_1\frac{q^{2}_{1\perp}
-2q_{1\perp}k_{2\perp}}{\tilde
t_2}-\frac{x_1q^{\alpha}_{1\perp}k^{\beta}_{2\perp}
-x_2(q_1-k_2)^{\alpha}_{\perp}q^{\beta}_{1\perp}}{x_2\tilde
t_2}
-\frac{q^2_{1\perp}(q_1-k_2)^{\alpha}_{\perp}k^{\beta}_{2\perp}
}{k^2_{2\perp}\tilde t_2}
 ~, \label{bsymmetric}
\end{equation}
looks simpler than $b^{\alpha\beta}(q_1;k_1,k_2)$ and
$b^{\beta\alpha}(q_1;k_2,k_1)$ taken separately. Note,
however, that $F_s(k_1, k_2)$ does not turn into zero at
the points $x_1=0$ and $x_2=0$, in contrast to
$F_b(k_1,k_2)$, so that the prescription (\ref{integral
+}) is necessary in Eq.~(\ref{symmetric kernel}) to
remove the ``Regge divergencies", i.e. the divergencies
at $x\equiv x_1=0$ and $x=1$.  We shall use the
decompositions
\[
 b^{\alpha\beta}(q_1;k_1,k_2)+b^{\beta\alpha}(q_1;k_2,k_1)=
b^{\alpha\beta}_0+b^{\alpha\beta}_1+b^{\alpha\beta}_2~,
\]
\begin{equation}
b^{\alpha\beta}(q^{\prime}_1;k_1,k_2)+b^{\beta\alpha}(q^{\prime}_1;k_2,k_1)=
b^{\prime~\alpha\beta}_0+b^{\prime~\alpha\beta}_1+b^{\prime~\alpha\beta}_2~,
\label{bs=sum b}
\end{equation}
with
\begin{equation}
 b^{\alpha\beta}_0=\frac{q^2_{1\perp}k^{\alpha}_{1\perp}k^{\beta}_{2\perp}}
{k^2_{1\perp}k^2_{2\perp}}~,\label{b0}
\end{equation}
\begin{equation}
 b^{\alpha\beta}_1=-\frac{1}{2}g^{\alpha\beta}_{\perp}x_2
\frac{q^2_{1\perp}-2q_{1\perp}k_{1\perp}}{\tilde
t_1}-\frac{x_2k^{\alpha}_{1\perp}q^{\beta}_{1\perp}
-x_1q^{\alpha}_{1\perp}(q_1-k_1)^{\beta}_{\perp}}{x_1\tilde
t_1}
-\frac{q^2_{1\perp}k^{\alpha}_{1\perp}(q_1-k_1)^{\beta}_{\perp}
}{k^2_{1\perp}\tilde t_1}~\label{b1}
\end{equation}
and
\begin{equation}
b^{\alpha\beta}_2=-\frac{1}{2}g^{\alpha\beta}_{\perp}x_1\frac{q^{2}_{1\perp}
-2q_{1\perp}k_{2\perp}}{\tilde
t_2}-\frac{x_1q^{\alpha}_{1\perp}k^{\beta}_{2\perp}
-x_2(q_1-k_2)^{\alpha}_{\perp}q^{\beta}_{1\perp}}{x_2\tilde
t_2}
-\frac{q^2_{1\perp}(q_1-k_2)^{\alpha}_{\perp}k^{\beta}_{2\perp}
}{k^2_{2\perp}\tilde t_2}
 ~. \label{b2}
\end{equation}
The tensor $b^{\alpha\beta}_0$ does not depend on $x$ at
all; as far as $b^{\alpha\beta}_{1,2}$ is concerned, it is easy to obtain
\begin{equation}
 b^{\alpha\beta}_1|_{x=0}=-\frac{k^{\alpha}_{1\perp}q^{\beta}_{1\perp}
}{ k_{1\perp}^2}~,\;\; b^{\alpha\beta}_2|_{x=0}=
\frac{(q_1-k_2)^{\alpha}_{\perp}q^{\beta}_{1\perp}}{(q_1-k_2)_{\perp}^2}
-\frac{q^2_{1\perp}(q_1-k_2)^{\alpha}_{\perp}k^{\beta}_{2\perp}
}{k^2_{2\perp}(q_1-k_2)_{\perp}^2}~, \label{x=0}
\end{equation}
\begin{equation}
 b^{\alpha\beta}_1|_{x=1}=
\frac{q^{\alpha}_{1\perp}(q_1-k_1)^{\beta}_{\perp}}{(q_1-k_1)_{\perp}^2}
-\frac{q^2_{1\perp}k^{\alpha}_{1\perp}(q_1-k_1)^{\beta}_{\perp}
}{k^2_{1\perp}(q_1-k_1)_{\perp}^2}~,\;\;b^{\alpha\beta}_2|_{x=1}=
-\frac{q^{\alpha}_{1\perp}k^{\beta}_{2\perp} }{
k_{2\perp}^2}~. \label{x=1}
\end{equation}
Note also that at fixed $x$ the tensors $b^{\alpha\beta}_i$
have infrared singularities. The singularities of
$b^{\alpha\beta}_0$ are evident from Eq.~(\ref{b0}). As for
$b^{\alpha\beta}_i$ with $i=1,2$, they are singular at
$k_{i\perp}$, where
\begin{equation}
 b^{\alpha\beta}_1|_{k_{1\perp}\rightarrow 0}=
-\frac{k^{\alpha}_{1\perp}q^{\beta}_{1\perp}
}{k^2_{1\perp}}~,\;\;\;\;b^{\alpha\beta}_2|_{k_{2\perp}\rightarrow
0}= -\frac{q^{\alpha}_{1\perp}k^{\beta}_{2\perp}
}{k^2_{2\perp}}~. \label{b12infrared}
\end{equation}
Accordingly to the composition (\ref{bs=sum b}), we present $F_s(k_1,
k_2)$ in the form
\begin{equation}
F_s(k_1, k_2)=(A_0+A_1+A_{2}+A_{3})+(k_1\leftrightarrow
k_2)~. \label{As=sumAi}
\end{equation}
Note that, since the total convolution (\ref{As=sumAi})
includes the terms obtained by the substitution
$(k_1\leftrightarrow k_2)$ from $A_{i}$, the choice of
$A_{i}$ is not unique. We get (in the following in this
Section only transverse momenta are used and we omit the
$\perp$ sign; pay attention, however, that the Minkowski
metric is used)
\begin{equation} A_0=
\frac{1}{2}b^{\alpha\beta}_0b^{\prime}_{0\;\alpha\beta}~=\frac{q^2_{1}q^{\prime
2}_{1}}{2k^2_1k^2_2}~, \label{A0}
\end{equation}
\[
A_1= b^{\alpha\beta}_1b^{\prime}_{0\;\alpha\beta}+
 b^{\alpha\beta}_0b^{\prime}_{1\;\alpha\beta}~
\]
\begin{equation}
=-\frac{q^{\prime 2}_{1}}{k^2_1k^2_2\tilde
t_1}\left[\frac{x_2}{2}(k_1k_2)(q^2_{1}-2q_{1}k_{1})+
\frac{x_2}{x_1}k^2_{1}(q_{1}k_2)+(q^2_{1}
-(q_{1}k_1))(k_2(q_1-k_1)) \right]+(q_1\leftrightarrow
q^{\prime}_1)~, \label{A1}
\end{equation}
\[
A_{2}=b^{\alpha\beta}_1b^{\prime}_{1\;\alpha\beta}~=\frac{1}{2\tilde
t_1\tilde t^{\prime}_1}\left[
\frac{(D-2)}{4}x^2_2(q^2_{1}-2q_{1}k_{1})(q^{\prime
2}_{1}-2q^{\prime~}_{1} k_{1})+x_2(q^{\prime 2}_{1}
-2q^{\prime}_{1}k_{1})\right.
\]
\[
\left.\times\left((q_{1}
k_{1})(\frac{1}{x_1}+\frac{q^{2}_{1}}{k_1^2})-2q_1^2\right)
+\frac{x_2^2}{x_1^2}k_1^2(q_1q^{\prime~}_{1})+
(q_1q^{\prime~}_{1})((q_1-k_1)(q^{\prime}_{1}-k_1))
-2\frac{x_2}{x_1}(k_1q_{1})(q^{\prime}_1(q_{1}-k_1))\right.
\]
\begin{equation}
\left.+\frac{q^{\prime
2}_1}{k_1^2}\left(2\frac{x_2}{x_1}k_1^2(q_1(q^{\prime~}_{1}
-k_{1}))+(q^{2}_1-2(q_1k_1))((q_{1}
-k_{1})(q^{\prime~}_{1} -k_{1}))
\right)\right]+(q_i\leftrightarrow q^{\prime}_i)~,
\label{A11}
\end{equation}
As for $A_{3}$,  its definition is not so simple.
Actually $A_{3}$ is constructed from
$b^{\alpha\beta}_1b^{\prime}_{2\;\alpha\beta}$. Note that
this convolution is invariant with respect to the
simultaneous substitution $k_1\leftrightarrow
k_2 $ and $q_1\leftrightarrow q^{\prime}_1$. As for the two
terms in $b^{\alpha\beta}_1b^{\prime}_{2\;\alpha\beta}$,
which are obtained from each other by this substitution,
we have taken  one of them in an unchanged form, whereas
in the other we have performed the substitution
$k_1\leftrightarrow k_2$, so that it can be obtained from
the first term by the substitution $q_1\leftrightarrow
q^{\prime}_1$. Of course, this procedure is not unique.
We define $A_{3}$ as
\[
A_{3}= \frac{1}{2\tilde t_1\tilde
t^{\prime}_2}\left[\frac{ (D-2)}{4}
x_1x_2(q^2_{1}-2q_{1}k_{1})(q^{\prime
2}_{1}-2q^{\prime~}_{1} k_{2})+(q^{\prime 2}_{1}
-2q^{\prime}_{1}k_{2})\left((q_{1}
k_{1})(1+\frac{x_1q^{2}_{1}}{k_1^2})-2x_1q^2_{1}\right)\right.
\]
\[
\left.+(q_{1}k_{2})(q^{\prime}_{1}k_{1})+
(q_1(q^{\prime}_{1}-k_{2}))(q^{\prime}_{1}(q_1-k_{1}))
-\frac{2x_2}{x_1}(q_1q^{\prime}_{1})(k_1(q^{\prime}_1-k_{2}))
\right.
\]
\[
\left.+2q^{\prime
2}_{1}\frac{x_2(q_{1}k_{2})(q^{\prime~}_{1}
k_{1}-k_2k_1)-x_1(q_{1}
k_{2}-k_1k_2)(q_{1}q^{\prime}_{1}-q_{1}k_{2})}{x_1k_2^2}\right.
\]
\begin{equation}
\left.+\frac{q^{2}_{1}q^{\prime 2}_{1}}{k_1^2k_2^2}(q_{1}
k_{2}-k_1k_2)(q^{\prime~}_{1}
k_{1}-k_2k_1)\right]+(q_i\leftrightarrow q^{\prime}_i)~.
\label{A12}
\end{equation}
From Eqs.~(\ref{symmetric kernel}) and (\ref{As=sumAi}) it
follows that we can present ${\cal K}^{({s})}_{GG}(\vec
q_1,\vec q_2;\vec q)$ as
\begin{equation}
{\cal K}^{({s})}_{GG}(\vec q_1,\vec q_2;\vec q)=
\frac{4g^4N_c^2}{(2\pi)^{D-1}}
\frac{\Gamma(1-\epsilon)}{(4\pi)^{2+\epsilon}}\hat{{\cal
S}}({\cal J}_0+{\cal J}_1+{\cal J}_{2}+{\cal
J}_{3})~,\label{expansion of symmetric kernel}
\end{equation}
where
\begin{equation}
{\cal J}_i= \int_{0}^{1}{dx}\int\frac{d^{2+2\epsilon}k_1}
{\pi^{1+\epsilon}\Gamma(1-\epsilon)}\left(\frac{A_i
+(k_1\leftrightarrow k_2)}{x(1-x)}\right)_+~.
\label{integrals Ji}
\end{equation}
Since $A_0$ does not depend on $x$, the integral ${\cal
J}_{0}$ evidently is equal to zero according to the
definition (\ref{integral +}). The integrals ${\cal
J}_{i}$ with $i\neq 0$ will be  discussed below. Here we
note that ${\cal K}^{({s})}_{GG}(\vec q_1,\vec q_2;\vec
q)$ is finite in the limit $\epsilon \rightarrow 0$.
Indeed, let us consider the terms in $A_i$  having
non-integrable infrared singularities in the limit
$\epsilon\rightarrow 0$ . These terms can be easily
obtained using Eqs.~(\ref{b0}) and (\ref{b12infrared}),
as well as the explicit expressions
(\ref{A1})-(\ref{A12}):
\begin{equation}
A_1|_{sing}= -\frac{q^{\prime
2}_{1}(q_{1}k_2)}{k^2_1k^2_2}+(q_1\leftrightarrow
q^{\prime}_1)~,\;\;\;A_{2}|_{sing}=
\frac{(q_1q_1^{\prime})}{2k_1^2}+(q_1\leftrightarrow
q^{\prime}_1)~, \;\;\; A_{3}|_{sing}=0~.~\label{Asing}
\end{equation}
We see that the infrared singular parts of $A_i$ do not
depend on $x$. On the other hand for $A_i|_{x=0}$ and
$A_i|_{x=1}$ we find (it can be done using
Eqs.~(\ref{b0}), (\ref{x=0}) and (\ref{x=1}) as well as
Eqs.~(\ref{A1})-(\ref{A12}) )
\[
A_1|_{x=0}= -\frac{q^{\prime
2}_{1}(q_{1}k_2)}{k^2_1k^2_2}+(q_i\leftrightarrow
q^{\prime}_i)~,\;\;
\]
\begin{equation}
A_1|_{x=1}=-\frac{q^{\prime
2}_{1}}{k^2_1k^2_2(q_1-k_1)^2}(q^2_{1}
-(q_{1}k_1))(k_2(q_1-k_1)) +(q_i\leftrightarrow
q^{\prime}_i)~, \label{A1as}
\end{equation}
\[
A_{2}|_{x=0}=\frac{(q_1q_1^{\prime})}{2k_1^2}+(q_i\leftrightarrow
q^{\prime}_i)~,\;\;
\]
\begin{equation}
A_{2}|_{x=1}=\frac{(q_{1}-k_1)(q^{\prime~}_{1}-k_1)}
{2(q_{1}-k_1)^2(q^{\prime~}_{1}-k_1)^2}\left[(q_{1}q^{\prime~}_{1})
+\frac{q^{\prime
2}_{1}}{k_1^2}\left(q_1^2-2(k_1q_1)\right)\right]+(q_i\leftrightarrow
q^{\prime}_i)~, \label{A11as}
\end{equation}
\begin{equation}
A_{3}|_{x=0}=
-\frac{(k_1(q^{\prime}_1-k_{2}))}{k_1^2(q_1^{\prime}-k_2)^2}
\left[ (q_1q^{\prime}_{1})-q^{\prime
2}_{1}\frac{(q_{1}k_{2})}
{k_2^2}\right]+(q_i\leftrightarrow
q^{\prime}_i),\;\;\;A_{3}|_{x=1}=0~. \label{A12as}
\end{equation}
Comparing these expressions with the results (\ref{Asing}), we see
that they do not contain new (i.e. different from $A_i|_{sing}$)
non-integrable infrared singularities in the limit $\epsilon\rightarrow 0$.
Therefore such singularities are absent in $\left(A_i/{[x(1-x)]}\right)_+$.

\subsection{Calculation of ${\cal J}_1$}

In order to calculate the integral ${\cal J}_1$ it is suitable
to first  integrate over $x$ and then over $k_1$. Using
the invariance of the integration measure with respect to the
exchange $k_1\leftrightarrow k_2$, after the first integration
(here and below we omit in integrands terms giving zero after the
subsequent integration) we obtain
\[
{\cal J}_1=\frac{\ q_1^{\prime
~2}}{2}\int\frac{d^{2+2\epsilon}k_1}
{\pi^{1+\epsilon}\Gamma(1-\epsilon)}\frac{1}{\
k_2^{2}}\ln\left(\frac{(\ q_1-\ k_1)^2}{\ k_1^{2}}\right)
\]
\begin{equation}
\times\left(\frac{\ q_1^{2}\ q_2^{2}}{\ k_1^{2}(\ q_1-\
k_1)^2}-\frac{\ q_2^{2}}{(\ q_1-\ k_1)^2}-\frac{\
q_2^{2}+2q_2k_1}{\
k_1^{2}}\right)\;+\;(q_i\leftrightarrow q^{\prime}_i)~.
\label{J1 over x}
\end{equation}
Note that the singularities of separate terms in the
integrand at $\ k_1=0, \;\;\ k_2=0$ and $\ q_1- k_1=0$
cancel each other.

Details of the calculation of this integral are given in
Appendix B. At arbitrary $\epsilon$ we find
\[
{\cal J}_1=\frac{\ q_1^{\prime ~2}}{2} \int_0^1dx
\int_0^1
dyy^{\epsilon-1}\left[\frac{(1-\epsilon)q_{1}^2q_2^2}
{\left(-(k^2(1-x)+q_2^2 x)(1-y)-q_1^2x(1-x)y\right)^{2
-\epsilon}}\ln\left(\frac{x}{1-x}\right)\right.
\]
\begin{equation}
\left.
+\left(\frac{(q_1^2-k^2)(1-x)(1-y(1-x))-q_2^2(x-y(1-x^2))}
{\left(-(k^2(1-x)+q_2^2x)(1-y)-q_1^2x(1-x)y\right)^{1
-\epsilon}x(1-x)}\right)_+\right]\;+\;(q_i\leftrightarrow
q^{\prime}_i)~. \label{J1 at arbitrary d}
\end{equation}
The integral cannot be expressed in terms of elementary
functions not only at arbitrary $D$, but even in the
limit $\epsilon\rightarrow 0$. In this limit we have
\begin{equation}
{\cal J}_1=\frac{\ q_1^{\prime ~2}}{2}\left(
\frac{(k^2-q_{1}^2-q_{2}^2)^2-4q_{1}^2q_{2}^2}{2k^2}I(k^2,
q_2^2, q_1^2)
+\frac{k^2+q_{2}^2-q_{1}^2}{2k^2}\ln\left(\frac{k^2}{q_2^2}\right)
\ln\left(\frac{q_1^2}{q_2^2}\right)\right)\;+\;(q_i\leftrightarrow
q^{\prime}_i)~,\label{J1 at d=4}
\end{equation}
where
\begin{equation}
I(a, b, c)= \int_0^1\frac{dx}{ a(1-x)+b x-c x(1-x)}\ln
\left( \frac{ a(1-x)+ b x}{ c x(1-x)}\right).
\label{I(p,q,r)}
\end{equation}
Note that the integral $I(a,b,c)$ is invariant with
respect to any permutation of its arguments, as it can be seen
from the representation~\cite{FP2002}
\begin{equation}
I(a,b,c)=\int_0^1\int_0^1\int_0^1\frac{dx_1 dx_2
dx_3\delta(1-x_1-x_2-x_3)}{(ax_1+ bx_2+ c
x_3)(x_1x_2+x_1x_3+x_2x_3)} ~.\label{symmetric integral}
\end{equation}
In particular, $I(k^2, q_2^2, q_1^2)$ does not change
under the substitution $q_1\leftrightarrow -q_2$.

\subsection{Calculation of ${\cal J}_2$}
The order of integration used for the calculation of
${\cal J}_1$  (first over $x$ and then over $k_1$) is
suitable for the calculation of  ${\cal J}_2$ as well.
Details of the integration are given in Appendix C. The
result of the integration over $x$ can be presented as
\[
{\cal J}_2=
\int\frac{d^{2+2\epsilon}k_1}{\pi^{1+\epsilon}\Gamma(1-\epsilon)}
\left\{
\left[\frac{1}{(q_1-k_1)^2-k_1^2}\left((1+\epsilon)k_1^2-q_1^2
\left(2-\frac{(k_1q_1)}{k_1^2}\right)\right)+\frac{(q_1-k_1)
(q_1^{\prime}-k_1)}{(q_1-k_1)^2}\right.\right.
\]
\[
\left.\left.\times\left(\frac{q_1^2}{k_1^2}-\frac{q^2}{2(q_1^{\prime}-k_1)^2}\right)-
\frac{(q_1q_1^{\prime})}{k_1^2}\right]
\ln\left(\frac{(q_1-k_1)^2}{k_1^2}\right)+
\left[\left(\frac{q^2}{2}\left(1+\frac{(q_1-k_1)
(q_1^{\prime}-k_1)}{(q_1-k_1)^2}\right)
\right)\right.\right.
\]
\begin{equation}
\left.\left.-\frac{1+\epsilon}{2}(q_1-k_1)^2\right]\frac{1}
{(q_1-k_1)^2-(q_1^{\prime}-k_1)^2}\ln\left(\frac{(q_1-k_1)^2}{(q_1^{\prime}-k_1)^2}\right)
\right\}+(q_1\leftrightarrow q^{\prime}_1)~. \label{J2
integrated over x}
\end{equation}
At arbitrary $\epsilon$ the integration in Eq.~(\ref{J2 integrated
over x}) gives
\[
{\cal J}_2=
\frac{\Gamma^2(1+\epsilon)}{\epsilon\Gamma(1+2\epsilon)}
\left((-q_1^2)^{1+\epsilon}+(-q_1^{\prime\:2})^{1+\epsilon}-
(-q^2)^{1+\epsilon}\right)
\]
\begin{equation}
\times \left(\frac{11+7\epsilon}{2(1+2\epsilon)
(3+2\epsilon)}-\psi(1+\epsilon)+\psi(1+2\epsilon)\right)
- \frac{q^2}{2}I_+(q_1, q_1^\prime)~, \label{J2 at
arbitrary d}
\end{equation}
where
\begin{equation}
I_+(q_1,
q_1^\prime)=-\int\frac{d^{2+2\epsilon}k_1}{\pi^{1+\epsilon}
\Gamma(1-\epsilon)}\frac{(q_1-k_1)(q_1^{\prime}-k_1)}{(q_1-k_1)^2
(q_1^{\prime}-k_1)^2}
\ln\left(\frac{(q_1-k_1)^2(q_1^{\prime}-k_1)^2}{(k_1^2)^2}\right)
~. \label{I_+}
\end{equation}
This last integral cannot be expressed in terms of
elementary functions at arbitrary $\epsilon$. In the
limit $\epsilon\rightarrow 0$, instead, at
fixed nonzero $q$ it becomes
\begin{equation}
I_+(q_1, q_1^\prime)=-\ln\left(\frac{q_1^2}{q^2}\right)
\ln\left(\frac{q_1^{\prime\:2}}{q^2}\right)~.
\label{limit of I_+}
\end{equation}
Note that for $I_+(q_1, q_1^\prime)$ the limits
$\epsilon\rightarrow 0$ and $q\rightarrow 0$ are
noninterchangeable. However, it does not matter for ${\cal
J}_2$, where $I_+(q_1, q_1^\prime)$ enters with the
coefficient $\sim q^2$. At $\epsilon\rightarrow 0$ we
obtain
\[
{\cal
J}_2=({q_1^2+q_1^{\prime\:2}-q^2})\left(-\frac{11}{6\epsilon}
+\frac{67}{18}-\zeta(2)\right)+\frac{11}{6}\left(q_1^2\ln(-q_1^2)+
q_1^{\prime\:2}\ln(-q_1^{\prime\:2})-q^2\ln(-q^2)\right)
\]
\begin{equation}
-\frac{q^2}{2} \ln\left(\frac{q_1^2}{q^2}\right)
\ln\left(\frac{q_1^{\prime\:2}}{q^2}\right)~. \label{J2
at d=4}
\end{equation}
Appearance of the pole at $\epsilon =0$ in ${\cal J}_2$
does not contradict the statement that non-integrable
infrared singularities in the limit $\epsilon\rightarrow
0$ are absent in $\left(A_i/{[x(1-x)]}\right)_+$. The
pole term comes from the ultraviolet region. Indeed, one
can see from the expression (\ref{A11}) for $A_2$ that,
after averaging over the azimuthal angle at $D=4$ in the
limit of large $|k_1|$, we have
\begin{equation}
\left(\frac{{\overline A_2}}{x(1-x)}\right)_+\simeq
\frac{q_1q_1^{\prime}}{k_1^2}(x(1-x)-2)~.
\label{ultraviolet limit of A2}
\end{equation}
The pole term in Eq.~(\ref{J2 at d=4}) is just the
doubled result of the integration of the expression
(\ref{ultraviolet limit of A2}) over
$dxd^{2+2\epsilon}k_1/\pi$.  Evidently, the ultraviolet
divergency is artificial and appears to be the result of
the separation of $F_s(k_1, k_2)$ as shown in
Eq.~(\ref{As=sumAi}). Indeed, as it can be easily proved
from formulas (\ref{A11}) and (\ref{A12}), the terms
leading to such divergencies cancel in the sum
$A_2+A_3+(k_1\leftrightarrow k_2)$. This means that the
pole term in ${\cal J}_{2}$ is cancelled by an analogous
term in ${\cal J}_{3}$ (see below).

\subsection{Calculation of ${\cal J}_{3}$}

The integral ${\cal J}_{3}$ is much more complicated than
the preceding ones. As a consequence, the trick of
integrating first over $x$, applied
before, cannot be used in the calculation of ${\cal
J}_{3}$, because it leads to terms with the denominators
containing a third power of $k_1$. Such terms cannot be
integrated over $k_1$ by known methods. This complexity
is connected with non-planarity of diagrams represented
by ${\cal J}_{3}$, which is seen from the denominator
$\tilde t_1 \tilde t_2$ related to the cross-box diagram.
The complexity of contributions of the cross-box diagrams
is well known since the calculation of the non-forward
kernel for the QED Pomeron \cite{Non-forward Pomeron in
QED} which was found only in the form of  two-dimensional
integral. In QCD the situation is greatly worse because
of the existence of cross-pentagon and cross-hexagon
diagrams in addition to QED-type cross-box diagrams.  It
requires the use of additional Feynman parameters. At
arbitrary $D$ no integration over these parameters at all
can  be done in elementary functions. It occurs, however,
that in the limit $\epsilon \rightarrow 0$ the
integration over additional Feynman parameters can be
performed, so that the result can be written as
two-dimensional integral, as well as in QED. Details of
the calculation are given in Appendix D. The result is
\[
{\cal J}_3=
\frac{11}{12}({q_1^2+q_1^{\prime\:2}-q^2})\left(
\frac{1}{\epsilon}+\ln(-k^2)+1\right)+J(q_1, q_2;q)+
(q_i\leftrightarrow q^{\prime}_i)
\]
with
\[
J(q_1, q_2;q)=\int_{0}^{1}dx\int_0^1
dz\left\{q_1q_1^{\prime}\left((2-x_1x_2)\ln
\left(\frac{Q^{2}}{-k^2}\right) -\frac{2}{x_1}\ln
\left(\frac{Q^{2}}{Q_0^{2}}\right)\right)\right.
\]
\[
\left.+\frac{1}{2Q^{2}}
x_1x_2(q^2_{1}-2q_{1}r_{1})(q^{\prime
2}_{1}-2q^{\prime~}_{1}
r_{2})-\frac{2}{x_1}\left[\left(x_2q_1q^{\prime~}_{1}(r_1(q^{\prime}_{1}
-r_2))-q^{\prime
2}_{1}q_1r_2\right)\frac{1}{Q^{2}}\right.\right.
\]
\[
\left. + \left(z(1-z)q^{\prime 2}_{2}q_1q^{\prime~}_{1}
+q^{\prime
2}_{1}(zq_1k+(1-z)q_1q^{\prime}_{1})\right)\frac{1}{Q_0^{2}}
\right]
\]
\[
\left. +\frac{1}{Q^2}\left(q^{\prime
2}_{1}q_1\left(r_{1}-2q^{\prime}_{1}\right)
+4x_1q_1^2(q^{\prime}_{1}r_2)+q^{\prime}_{1}q_{1}(q^{\prime}_{1}
q_{1}-q^{\prime}_{1}r_{1}-q_{1}r_{2})+2(q^{\prime}_{1}r_{1})
(q_{1}r_2)-2(q^{\prime}_{1}r_{2})(q_{1}r_1)\right)
\right.
\]
\[
\left. +{q^{\prime
2}_{1}}\left[\left(q_1(x_2q^{\prime}_{1}+q_2)-\frac{x_2}{x_1}q_1
(q_1^{\prime}+k)\right)\frac{1}{r_2^2}\ln\left(\frac{Q^2}{\mu_2^2}\right)
+\frac{1}{x_1}q_1(q_1^{\prime}+k)\frac{1}{r_0^2}\ln\left(\frac{Q_0^2}
{\mu_0^2}\right) \right)\right.
\]
\[
\left.\left. -\frac{1}{\mu_2^2Q^2}
\left(2\frac{x_2}{x_1}(q_1r_2)q^{\prime}_{1}k+x_2(q^{\prime}_{1}
r_2)(q_2^2-k^2)+2(q_2r_2)q_1q\right)
+\frac{2}{\mu_0^2Q_0^2}\frac{1}{x_1}(q_1r_0)q^{\prime}_{1}k
\right.\right.
\]
\[
+\frac{1}{r_2^2}\left(\frac{1}{r_2^2}\ln\left(\frac{Q^2}{\mu_2^2}
\right)-\frac{1}{Q^2}\right)
\left(2\frac{x_2}{x_1}(q_1r_2)(q^{\prime}_{1}+k)r_2-2((x_2
q^{\prime}_{1}+q_2)r_2)q_1r_2\right)
\]
\[
-\frac{1}{r_0^2}\left(\frac{1}{r_0^2}\ln\left(\frac{Q_0^2}{\mu_0^2}
\right)-\frac{1}{Q_0^2}\right)
\left(2\frac{1}{x_1}(q_1r_0)(q^{\prime}_{1}+k)r_0\right)
\]
\[
+\frac{q_1^2}{d}\left(-(q_2k)(q_2^{\prime}k)\left(\frac{1}{k^2}
+\frac{Q^2}{d}{\cal
L}\right)-(q_2r_2)(q_2^{\prime}k)\left(\frac{1}{\mu_2^2}
-\frac{\mu_1^2}{d}{\cal
L}\right)-(q_2k)(q_2^{\prime}r_1)\left(\frac{1}{\mu_1^2}
-\frac{\mu_2^2}{d}{\cal L}\right)\right.
\]
\begin{equation}
\left.\left.\left.+(q_2r_2)(q_2^{\prime}r_1)\left(\frac{1}{Q^2}
+\frac{k^2}{d}{\cal
L}\right)+\frac{(q_2q_2^{\prime})}{2}{\cal
L}\right)\right]\right\}~. \label{J3 at d=4}
\end{equation}
Here
\[
r_1=zxq_1+(1-z)(xk-(1-x)q_2^{\prime}),\;\;r_2=z((1-x)k-xq_2)
+(1-z)(1-x)q_1^{\prime}; \;\; r_1+r_2=k,
\]
\[
Q^2=-x(1-x)(q_1^2 z+q_1^{\prime 2}(1-z))-z(1-z)(q_2^2
x+q_2^{\prime 2}(1-x)-q^2x(1-x)),\;\;\;
\mu_i^2=Q^2-r_i^2,
\]
\[
r_0=zk+(1-z)q_1^{\prime},\;\;\; Q_0^2=-z(1-z)q^{\prime
2}_2,\;\;\; \mu_0^2=-zk^2-(1-z)q^{\prime 2}_1,
\]
\[
d=\mu_1^2\mu_2^2+k^2Q^2=z(1-z)x(1-x)\left((k^2-q_1^2-q_2^{\prime
2})(k^2-q_1^{\prime
2}-q_2^2)+k^2q^2\right)+q_1^2q_2^2xz(x+z-1)
\]
\begin{equation}\label{definition d and L}
+q_1^{\prime 2}q_2^{\prime 2}(1-x)(1-z)(1-x-z),\;\;\;
{\cal L}=\ln\left(\frac{\mu_1^2\mu_2^2}{-k^2Q^2}\right).
\end{equation}

\Section{Non-forward kernel}

In pure gluodynamics, which is considered here, the part
${\cal K}^{\left({\cal R}\right)}_r$ of the BFKL kernel
(\ref{kernel=virtual+real}), related to the production of
real particles,  for any representation ${\cal R}$ is
given by the sum of one-gluon ${\cal K}^{\left( {\cal
R}\right)}_G$ and two-gluon ${\cal K}^{\left( {\cal
R}\right)}_{GG}$ contributions.  Using for the last of
them the decomposition (\ref{kernel through symmetric
kernel}) we have
\begin{equation}
{\cal K}^{\left( {\cal R}\right)}_r={\cal K}^{\left(
{\cal R}\right) }_G+4(a_R-b_R){\cal K}^{({8})}_{GG} +b_R
{\cal K}^{({s})}_{GG}~,\label{KR through K8 and KS}
\end{equation}
where the colour group coefficients $a_R$ and $b_R$ are
defined in Eqs.~(\ref{coefficients a and b}) and the
one-gluon contribution ${\cal K}^{\left( {\cal R}\right)
}_G$ is given by Eq.~(\ref{one gluon contribution}) (for
arbitrary $D$ see Eq.~(4.10) in the last of
Refs.~\cite{vertex}). The two-gluon contribution for the
octet channel ${\cal K}^{({8})}_{GG}$ was calculated  in
the second of Refs.~\cite{gluon octet kernel} (see there
Eqs.~(61) and (63) for arbitrary $D$ and for $D=4$,
respectively). The calculation of the ``symmetric"
contribution ${\cal K}^{({s})}_{GG}$ performed in this
paper solves the problem of finding the expression of the
non-forward BFKL kernel  for all possible colour states
in the $t$-channel. This contribution is determined by
Eq.~(\ref{expansion of symmetric kernel}), where ${\cal
J}_0=0$ and ${\cal J}_i$, for $i=1\div 3$, are given by
Eqs.~(\ref{J1 at arbitrary d}), (\ref{J2 at arbitrary
d}), (\ref{J3 at arbitrary d}) and (\ref{J1 at d=4}),
(\ref{J2 at d=4}), (\ref{J3 at d=4}) for arbitrary $D$
and for $D=4$ correspondingly. Note that everywhere in
these formulas the bare coupling constant $g$ is used.
The transition to the renormalized coupling $g_\mu $ in
the ${\overline{MS}}$ scheme takes place by means of
Eq.~(\ref{coupling renormalization}).

For the most important colour singlet case, using
$c_1=a_1=1$ and $c_8=b_1=1/2$, from Eqs.~(\ref{KR through K8 and KS})
and (\ref{one gluon contribution}) we obtain
\begin{equation}
{\cal K}^{\left( 1\right)}_r=2{\cal K}^{({8})}_r
+\frac{1}{2} {\cal K}^{({s})}_{GG}~.\label{K1 through K8
and KS}
\end{equation}
Because of the significance of this case let us consider it
in more detail.  We present the kernel ${\cal
K}^{\left( 1\right)}_r$ obtained from the above stated
sources in the limit $D=4+2\epsilon \rightarrow 4$ as
sum of two parts,
\begin{equation}
{\cal K}^{\left( 1\right)}_r={\cal K}^{sing}_r +{\cal
K}^{reg}~, \label{K1 through Ksing and Kreg}
\end{equation}
where the first, given by
\[
{\cal K}_r^{sing}(\vec q_1,\vec q_2; \vec{q})
=\frac{2\bar g_\mu ^2\mu^{-2\epsilon}}{\pi
^{1+{\epsilon}}\Gamma(1-\epsilon)}\left(
\frac{\vec{q}_1^{\:2} \vec{q}_2^{\:\prime\:2}+
\vec{q}_1^{\:\prime\:2} \vec{q}_2^{\:2}}{\vec k ^{\:2}}-
\vec{q}^{\:2}\right)\Biggl\{ 1+ \bar g_\mu
^2\biggl[\frac{11}{3\epsilon}
\]
\begin{equation}
+\left(\frac{\vec k^{\:2}}{\mu^2}\right)^\epsilon
\biggl\{-\frac{11}{3\epsilon}+\frac{67}{9}
-2\zeta(2)+\epsilon\left( -\frac{404}{27} +14\zeta(3)
+\frac{11}{3}\zeta(2)\right)\biggl\}\biggr]\Biggr\}\label{Ksingular}~,
\end{equation}
contains all singularities and the second, putting
$\epsilon=0$ and $\bar
 g_\mu^2=\alpha_s(\mu^2)N_c/(4\pi)$~, becomes
\[
{\cal K}_r^{reg}(\vec q_1,\vec q_2; \vec{q})
=\frac{\alpha_s^2(\mu^2)N^2_c}{16\pi^3} \Biggl[2(\vec
q_1^{\:2}+\vec q_2^{\:2}-\vec q^{\:2})
\left(\zeta(2)-\frac{50}{9}\right)
-\frac{11}{3}\left(\vec q_1^{\:2}\ln\left(\frac{\vec
q_1^{\:2}}{\vec k^{\:2}}\right)\right.
\]
\[
\left. +\vec q_2^{\:2}\ln\left(\frac{\vec q_2^{\:2}}{\vec
k^{\:2}}\right)-\vec q^{\:2}\ln\left(\frac{\vec
q_1^{\:2}\vec q_2^{\:2}}{\vec k^{\:4}}\right)-\frac{\vec
q_1^{\:2}\vec q_2^{\:\prime\:2}-\vec q_2^{\:2}\vec
q_1^{\:\prime\:2}}{\vec k^{\:2}}\ln\left(\frac{\vec
q_1^{\:2}}{\vec q_2^{\:2}}\right)\right)+\vec
q^{\,2}\left( \ln\left(\frac{\vec q_1^{\:2}}{\vec
q^{\:2}}\right)\ln\left(\frac{\vec q_1^{\:\prime 2}}{\vec
q^{\:2}}\right)\right.
\]
\[
\left.+\ln\left(\frac{\vec q_2^{\:2}}{\vec
q^{\:2}}\right)\ln\left(\frac{\vec q_2^{\:\prime 2}}{\vec
q^{\:2}}\right)+\frac{1}{2}\ln^2\left(\frac{\vec
q_1^{\:2}}{\vec
q_2^{\:2}}\right)\right)+\ln\left(\frac{\vec
q_1^{\:2}}{\vec q_2^{\:2}}\right)\left(\frac{\vec
q_1^{\:\prime 2}}{2}\ln\left(\frac{\vec q_2^{\:2}}{\vec
k^{\:2}}\right)-\frac{\vec q_2^{\:\prime
2}}{2}\ln\left(\frac{\vec q_1^{\:2}}{\vec
k^{\:2}}\right)\right.
\]
\[
\left.-\frac{\vec q_1^{\:2}\vec q_2^{\:\prime\:2}+\vec
q_2^{\:2}\vec q_1^{\:\prime\:2}}{2\vec
k^{\:2}}\ln\left(\frac{\vec q_1^{\:2}}{\vec
q_2^{\:2}}\right)+\frac{\vec q_1^{\:\prime\:2}(\vec
q_1^{\:2}-3\vec q_2^{\:2})}{2\vec
k^{\:2}}\ln\left(\frac{\vec k^{\:2}}{\vec
q_2^{\:2}}\right)+\frac{\vec q_2^{\:\prime\:2}(3\vec
q_1^{\:2}-\vec q_2^{\:2})}{2\vec
k^{\:2}}\ln\left(\frac{\vec k^{\:2}}{\vec
q_1^{\:2}}\right)\right)
\]
\[
+\left(\vec q^{\:2}(\vec k^{\:2}-\vec q_1^{\:2}-\vec
q_2^{\:2})+2\vec q_1^{\:2}\vec q_2^{\:2}-\frac{(\vec
q_1^{\:2}-\vec q_2^{\:2})(\vec q_1^{\:2}+\vec
q_2^{\:2})(\vec q_1^{\:\prime \:2}-\vec q_2^{\:\prime
\:2})}{2\vec k^{\:2}}+\vec q_1^{\:2}\vec q_1^{\:\prime
\:2}+\vec q_2^{\:2}\vec q_2^{\:\prime \:2}\right.
\]
\begin{equation}
\left.-\frac{\vec k^{\:2}}{2}(\vec q_1^{\:\prime
\:2}+\vec q_2^{\:\prime \:2})\right)I(\vec k^{\:2}, \vec
q_2^{\:2}, \vec q_1^{\:2})-2J(\vec q_1,\vec q_2;\vec
q)-2J(-\vec q_2, -\vec q_1;-\vec q)\Biggr] +\Biggl\{\vec q_i
\longleftrightarrow \vec q_i^{\:\prime}\Biggr\}~.
\label{Kregular}
\end{equation}
Here the functions  $I(k^2, q_2^2, q_1^2)$  and $J(q_1,
q_2;q)$ are defined in Eqs.~(\ref{I(p,q,r)}) and (\ref{J3
at d=4}) correspondingly.

All singularities of ${\cal K}^{\left( 1\right)}_r$ are
contained in the first part. We remind that ${\cal
K}^{({R})}_{G}$ and ${\cal K}^{({R})}_{GG}$ separately
contain first and second order poles at  $\epsilon =0$
(see Eq.~(\ref{one gluon contribution})). In the sum of
these contributions the pole terms cancel, so that at
fixed nonzero $\vec k^2$, when the term $\left({\vec
k^{\:2}}/{\mu^2}\right)^{\epsilon}$  in Eq.~(\ref{Ksingular})
can be expanded in $\epsilon$, the sum is finite at
$\epsilon =0$.  But the kernel (\ref{Ksingular}) is
singular at $\vec k^{\:2}=0$ so that, when it is
integrated over $q_2$, the region of so small $\vec
k^{\:2}$ values such that  $\epsilon |\ln \left({\vec
k^{\:2}}/{\mu^2}\right)|\sim 1$ does contribute.
Therefore the expansion of $\left({\vec
k^{\:2}}/{\mu^2}\right)^{\epsilon}$ is not  done in
Eq.~(\ref{Ksingular}). Moreover, the terms $\sim
\epsilon$ are taken into account in the coefficient of
the expression divergent at $\vec k^{\:2}=0$ in order to
save all contributions non-vanishing in the limit
$\epsilon\rightarrow 0$ after the integration.

As it was already discussed, the ``symmetric" part ${\cal
K}^{({s})}_{GG}$ of the kernel (\ref{KR through K8 and
KS}) is finite in the limit $\epsilon=0$. Moreover, it
does not give singularities at $\epsilon=0$ when the
kernel is used in the equation for the Green's function.
Indeed, the points $\vec q_2=0$ and $\vec
q_2^{\:\prime}=0$ do not give such singularities due to
the ``gauge invariance" properties (\ref{gauge invariance
of the kernel}), because these properties  are valid for
any representation. Its validity for  ${\cal
K}^{({s})}_{GG}$ can be checked explicitly using the
properties of $b^{\alpha\beta}$. It is not difficult also
to see that ${\cal K}^{({s})}_{GG}$ has not
non-integrable singularities
 in the limit $\epsilon=0$ at $\vec k=0$.

For the singlet case the infrared singularities of ${\cal
K}^{\left( {1}\right)}_r$ must be cancelled by the
singularities of the gluon trajectory after integration
of the total kernel with any nonsingular at $\vec k =0$
function.  The total BFKL kernel in the singlet case must
be free from singularities. It is not difficult to see
that it is the case, using the equality
\[
\omega(t) =-2\bar
g_\mu^2\left(\frac{1}{\epsilon}+\ln\left(\frac{-t}{\mu^2}\right)\right)
-\bar g_\mu^4
\left[\frac{11}{3}\left(\frac{1}{\epsilon^2}-\ln^2\left(\frac{-t}
{\mu^2}\right)\right)+\left(\frac{67}{9}-2\zeta(2)\right)\right.
\]
\begin{equation}
\left.\times
\left(\frac{1}{\epsilon}+2\ln\left(\frac{-t}{\mu^2}\right)\right)
-\frac{404}{27} +2\zeta(3) \right]~. \label{renormalized
NLO trajectory}
\end{equation}
It is convenient to represent the total kernel in such a
form that the cancellation of singularities between real
and virtual contributions becomes evident. For this
purpose let us first switch from the dimensional
regularization to the cut-off $\vec k^2 > \lambda^2$,
$\lambda\rightarrow 0$,  which is more convenient for
practical purposes.  With such regularization we can pass
to the limit $\epsilon\rightarrow 0$ in the real part of
the kernel, so that its singular part assumes the form
\[
{\cal K}_r^{sing}(\vec q_1,\vec q_2; \vec{q})\;
\rightarrow\; \theta((\vec q_1-\vec
q_2)^2-\lambda^2){\cal K}_{r}^{sing}(\vec q_1,\vec q_2;
\vec{q})|_{\epsilon=0}=\frac{\alpha_s(\mu^2)N_c}{2\pi^2
}\left( \frac{\vec{q}_1^{\:2} \vec{q}_2^{\:\prime\:2}+
\vec{q}_1^{\:\prime\:2} \vec{q}_2^{\:2}}{\vec k ^{\:2}}-
\vec{q}^{\:2}\right)
\]
\begin{equation}
\times \Biggl\{
1-\frac{\alpha_s(\mu)N_c}{4\pi}\left(\frac{11}{3}
\ln\left(\frac{\vec k^{\:2}}{\mu^2}\right)-\frac{67}{9}
+2\zeta(2)\right)\Biggr\}\theta((\vec q_1-\vec
q_2)^2-\lambda^2)~.
\end{equation}
The trajectory must be transformed in such a way that the cut-off
regularization gives the same result as the $\epsilon$
regularization:
\[
\omega (t)\;\rightarrow \;\omega_{\lambda}(t)=
\lim_{\epsilon\rightarrow
0}\biggl(\omega(t)+\frac{1}{2}\int
\frac{d^{2+\epsilon}q_2}{\vec q_2^{\:2}\vec
q_2^{\:\prime\:2}}{\cal K}_r^{(1)}(\vec q_1,\vec q_2;
\vec q)\theta((\vec q_1-\vec q_2)^2-\lambda^2)\biggr)
\]
\[
=-\frac{\alpha_s(\mu ^2)N_c}{2\pi}\biggl\{\ln
\left(\frac{-t}{\lambda^2}\right)-\frac{\alpha_s(\mu^2)N_c}{4\pi
}\left[
\frac{11}6\left(\ln^2\left(\frac{-t}{\mu^2}\right)\right.\right.
\]
\begin{equation}
\left.\left.-\ln^2\left(\frac{\lambda^2}{\mu^2}\right)\right)-\left(
\frac{67}9-\frac{\pi ^2}3\right)\ln
\left(\frac{-t}{\lambda^2}\right)+6\zeta
(3)\right]\biggr\}~.
\end{equation}
It is easy to check that the integral over $d^2q_2$ of the
total kernel (\ref{kernel=virtual+real}) with any function
non-singular at $\vec k=0$   is $\lambda$
-independent in the limit $\lambda\rightarrow 0$.
Moreover, it is equally easy to find a form of the kernel which
does not contain $\lambda$ . It is sufficient to find
a representation
\begin{equation}\label{omega as integral}
\omega_{\lambda}(-\vec q_1^{\;2})=\int d^2q_2
f_\omega(\vec q_1,\vec q_2)\theta((\vec q_1-\vec
q_2)^2-\lambda^2)
\end{equation}
with such a function $f_\omega$ that the singularity
non-integrable at $\vec k=\vec q_1-\vec q_2=\vec
q_1^{\:\prime}-\vec q_2^{\:\prime}=0$ is cancelled in the
``regularized virtual kernel"
\begin{equation}
{\cal K}^{reg}_{v}(\vec q_1,\vec q_2;
\vec{q})=f_\omega(\vec q_1,\vec q_2)+f_\omega(\vec
q_1^{\:\prime},\vec q_2^{\:\prime})+\frac{{\cal
K}_{r}^{sing}(\vec q_1,\vec q_2;
\vec{q})|_{\epsilon=0}}{\vec q_2^{\:2}\vec
q_2^{\:\prime\:2}}~.
\end{equation}
After that we can proceed to the limit $\lambda=0$, obtaining
\[
\left(\hat{{\cal K}}^{(1)}\Psi\right)(\vec q_1)=\int d^2
q_2\biggl\{{\cal K}^{reg}_{v}(\vec q_1,\vec q_2;
\vec{q})\Psi(\vec q_1)
\]
\begin{equation}
+\frac{{\cal K}_{r}^{sing}(\vec q_1,\vec q_2;
\vec{q})|_{\epsilon=0}}{\vec q_2^{\:2}\vec
q_2^{\:\prime\:2}}\left(\Psi(\vec q_2)-\Psi(\vec
q_1)\right)+\frac{{\cal K}_{r}^{reg}(\vec q_1,\vec q_2;
\vec{q})}{\vec q_2^{\:2}\vec q_2^{\:\prime\:2}}\Psi(\vec
q_2)\biggr\}~.
\end{equation}
Of course, the choice of the function $f_\omega$ contains a
large arbitrariness. One simple choice is
\[
f_\omega(\vec q_1,\vec
q_2)=-\frac{\alpha_s(\mu^2)N_c}{2\pi^2 }
\frac{\vec{q_1}^{\:2}}{\vec k ^{\:2}(\vec{q_1}^{\:2}+\vec
k ^{\:2})}
\]
\begin{equation}
\times \Biggl\{
1-\frac{\alpha_s(\mu)N_c}{4\pi}\left(\frac{11}{3}
\ln\left(\frac{\vec k^{\:2}}{\mu^2}\right)-\frac{67}{9}
+2\zeta(2)
+\left(6\zeta(3)-\frac{11}{3}\zeta(2)\right)\frac{\vec{k}^{\:2}}
{(\vec{q_1}^{\:2}+\vec k ^{\:2})} \right)\Biggr\}~.
\end{equation}

We have to say that the integral (\ref{J3 at d=4}) for $J(q_1, q_2;q)$
entering into all  kernels besides the octet ones definitely
is presented not in the best form. We have decided to present it
in such shape in order to give a possibility of further development
to people interested in this subject.
The results of our
efforts on simplification of the kernel and investigation of
its  properties will be presented in a subsequent paper.

\vspace{0.2cm} \noindent \underline{\bf Acknowledgments:}
V.S.~F. thanks  the Alexander von Humboldt foundation for the research
award,  the Dipartimento di
Fisica dell'Universit\`a della Calabria and the Istituto
Nazionale di Fisica Nucleare - gruppo collegato di
Cosenza for their warm hospitality while a part of this
work was done.

\vspace{2cm}

\Sectionn{Appendix A}
\renewcommand{\theequation}{A.\arabic{equation}}
For the colour group $SU(N_c)$ with $N_c=3$ the possible
representations ${\cal R}$ are $\underline 1,
\underline{8_a}, \underline{8_s}, \underline{10},
\underline{\overline{10}}, \underline{27}$. Corresponding
projection operators are
\begin{equation}
\langle bb^{\prime }|\hat {{\cal P}}_1|aa^{\prime
}\rangle =\frac{ \delta _{bb^{\prime }}\delta
_{aa^{\prime }}}{N_c^2-1}\ , \label{ A pomeron projector}
\end{equation}
\begin{equation}
\langle bb^{\prime }|\hat {{\cal P}}_{8_a}|aa^{\prime
}\rangle =\frac{f_{bb^{\prime}c}f_{aa^{\prime}c}}{N_c}~,
\label{A gluon projector}
\end{equation}
\begin{equation}
\langle bb^{\prime }|\hat {{\cal P}}_{8_s}|aa^{\prime
}\rangle = d_{bb^{\prime }c}d_{aa^{\prime
}c}\frac{N_c}{N_c^2-4}\ , \label{A symmetric projector}
\end{equation}
\begin{equation}
\langle bb^{\prime }|\hat {{\cal P}}_{10}|aa^{\prime
}\rangle = \frac{1}{4}\left[\delta_{ba} \delta_{b^{\prime
}a^{\prime }}-\delta_{ba^{\prime }}\delta_{b^{\prime }a}
-\frac{2}{N_c}f_{bb^{\prime}c}f_{aa^{\prime}c}+if_{ba^{\prime
}c} d_{b^{\prime }ac}+id_{ba^{\prime
}c}f_{b^{\prime}ac}\right]~, \label{decuplet projector}
\end{equation}
\begin{equation}
 \langle bb^{\prime }|\hat {{\cal P}}_{\overline{10}}|aa^{\prime
}\rangle = \frac{1}{4}\left[\delta_{ba} \delta_{b^{\prime
}a^{\prime }}-\delta_{ba^{\prime }}\delta_{b^{\prime }a}
-\frac{2}{N_c}f_{bb^{\prime}c}f_{aa^{\prime}c}-if_{ba^{\prime
}c} d_{b^{\prime }ac}-id_{ba^{\prime
}c}f_{b^{\prime}ac}\right]~, \label{anti-decuplet
projector}
\end{equation}
\[
 \langle bb^{\prime }|\hat {{\cal P}}_{27}|aa^{\prime
}\rangle =
\frac{1}{4}\left[\left(1+\frac{2}{N_c}\right)\left(\delta_{ba}
\delta_{b^{\prime }a^{\prime }}+\delta_{ba^{\prime
}}\delta_{b^{\prime
}a}\right)-\frac{2(N_c+2)}{N_c(N_c+1)}\delta_{bb^{\prime
}}\delta_{aa^{\prime }}\right.
\]
\begin{equation}
\left.-\left(1+\frac{2}{N_c+2}\right)
d_{bb^{\prime}c}d_{aa^{\prime}c}+d_{bac} d_{b^{\prime
}a^{\prime }c}+d_{b^{\prime }ac}d_{ba^{\prime
}c}\right]~. \label{27 projector}
\end{equation}
Here  $f_{abc}$ and $d_{abc}$ are defined by the relation
\begin{equation}
t^at^b=(d_{abc}+if_{abc})\frac{t^c}{2}+\delta_{ab}\frac{I}{2N_c},
\label{definition f and d}
\end{equation}
where $t^a$ are the group generators in the fundamental
representation, normalized by the requirement
$tr(t^at^b)={\delta_{ab}}/{2}$ and  $I$ is the identity
matrix.

For generality, we do not put here $N_c=3$, so that above
expressions are valid for the $SU(N_c)$ group with
arbitrary $N_c$. Corresponding representations in this
case have dimensions
\begin{equation}
n_1=1\ ~,\ \ \ \, n_{8_a}=n_{8_s}=N_c^2-1 ~, \ \ \ \,
n_{10}=n_{\overline{10}}=\frac{(N^2_c-4)(N_c^2-1)}{4}~, \
\ \ \, n_{{27}}=\frac{(N_c+3)N_c^2(N_c-1)}{4}~.
\end{equation}
However, at $N_c>3$ there is an additional representation
with  dimension
\begin{equation}
n_{N_c>3}=\frac{(N_c+1)N_c^2(N_c-3)}{4}
\end{equation}
and  projection operator
\[
 \langle bb^{\prime }|\hat {{\cal P}}_{N_c>3}|aa^{\prime
}\rangle =
\frac{1}{4}\left[\left(1-\frac{2}{N_c}\right)\left(\delta_{ba}
\delta_{b^{\prime }a^{\prime }}+\delta_{ba^{\prime
}}\delta_{b^{\prime
}a}\right)+\frac{2(N_c-2)}{N_c(N_c-1)}\delta_{bb^{\prime
}}\delta_{aa^{\prime }}\right.
\]
\begin{equation}
\left.+\left(1-\frac{2}{N_c-2}\right)
d_{bb^{\prime}c}d_{aa^{\prime}c}-d_{bac} d_{b^{\prime
}a^{\prime }c}- d_{ba^{\prime
}c}d_{b^{\prime}ac}\right]~. \label{extra projector}
\end{equation}
In $SU(3)$ this projection operator turns into zero due
to the equality
\begin{equation}
d_{bb^{\prime}c}d_{aa^{\prime}c}+d_{ba^{\prime }c}
d_{b^{\prime }ac}+d_{bac}d_{a^{\prime
}b^{\prime}c}=\frac{1}{3}\left(\delta_{bb^{\prime}}\delta_{aa^{\prime}}
+\delta_{ba^{\prime }} \delta_{b^{\prime
}a}+\delta_{ba}\delta_{a^{\prime }b^{\prime}}\right)~,
\label{identity}
\end{equation}
which holds at $N_c=3$. The following useful identities
with
\begin{equation}\label{F and D in terms of f and d}
T^a_{bc}=-if_{abc}, \;\;D^a_{bc}=d_{abc}, \;\;\left[F^a,
F^b\right]=if_{abc}F^c, \;\;\left[F^a,
D^b\right]=if_{abc}D^c
\end{equation}
are valid  at arbitrary $N_c$:
\[
Tr\left(T^a\right)=Tr\left(D^a\right)=Tr\left(T^aD^b\right)=0,\;\;
Tr\left(T^aT^b\right)=N_c\delta^{ab},\,\,
Tr\left(D^aD^b\right)=\frac{N^2_c-4}{N_c}\delta^{ab},\,\,
\]
\[
Tr\left(T^aT^bT^c\right)=i\frac{N_c}{2}f_{abc},\,\,
Tr\left(T^aT^bD^c\right)=\frac{N_c}{2}d_{abc},\,\,
\]
\[
Tr\left(D^aD^bT^c\right)=i\frac{N^2_c-4}{2N_c}f_{abc},\,\,
Tr\left(D^aD^bD^c\right)=\frac{N^2_c-12}{2N_c}d_{abc},\,\,
\]
\[
Tr\left(T^aT^bT^cT^d\right) =\delta_{ad}
\delta_{bc}+\frac{1}{2}\left(\delta_{ab}
\delta_{cd}+\delta_{ac}\delta_{bd}\right)+
\frac{N_c}{4}\left(f_{adi}f_{bci}+
d_{adi}d_{bci}\right),\;\;
\]
\[
Tr\left(T^aT^bT^cD^d\right)
=i\frac{N_c}{4}\left(d_{adi}f_{bci}-
f_{adi}d_{bci}\right),\;\;
\]
\[
Tr\left(T^aT^bD^cD^d\right) =\frac{1}{2}\left(\delta_{ab}
\delta_{cd}-\delta_{ac}\delta_{bd}\right)+
\frac{N^2_c-8}{4N_c}f_{adi}f_{bci}+
\frac{N_c}{4}d_{adi}d_{bci},\;\;
\]
\[
Tr\left(T^aD^bT^cD^d\right)
=-\frac{1}{2}\left(\delta_{ab}
\delta_{cd}-\delta_{ac}\delta_{bd}\right)+
\frac{N_c}{4}\left(f_{adi}f_{bci}+
d_{adi}d_{bci}\right),\;\;
\]
\[
Tr\left(T^aD^bD^cD^d\right)
=i\frac{2}{N_c}f_{adi}d_{bci}+i\frac{N^2_c-8}{4N_c}f_{abi}d_{cdi}+
i\frac{N_c}{2}d_{abi}f_{cdi},\;\;
\]
\[
Tr\left(D^aD^bD^cD^d\right)
=\frac{N^2_c-4}{N^2_c}\delta_{ad}\delta_{bc}+\frac{1}{2}\delta_{ac}
\delta_{bd}+\frac{N^2_c-8}{2N^2_c}\delta_{ab}\delta_{cd}
\]
\[
+\frac{N_c}{4}f_{adi}f_{bci}+\frac{N^2_c-16}{4N_c}d_{adi}d_{bci}
-\frac{4}{N_c}d_{abi}d_{cdi},\;\;
\]
\begin{equation}\label{colour group algebra}
f_{adi}f_{bci}+ d_{adi}d_{bci}-f_{abi}f_{cdi}-
d_{abi}d_{cdi}+\frac{2}{N_c}\left(\delta_{ad}
\delta_{bc}-\delta_{ab}\delta_{cd}\right)=0.\;\;
\end{equation}
These equalities (\ref{colour group algebra}) can be
derived from the relation (\ref{definition f and d}) and
the completeness of the matrices $t^a$ and $I$. The
properties of the projection operators
\begin{equation}\label{properties of projections operators}
\hat {\cal P }_i\hat {\cal P }_j=\delta_{ij}\hat {\cal P
}_i, \;\;\;\sum_i \hat {\cal P }_i =I
\end{equation}
can be easily obtained  with the help of these
equalities, as well as the coefficients $c_{\cal R}$:
\begin{equation}
 c_1=1\,,\,\,\,c_{8_a}=c_{8_a}=\frac{1}{2}\,, \,\, c_{10}=
 c_{\overline{10}}
 =0\,, \,\,c_{27}=-c_{N_c>3}=-\frac{1}{4N_c}~.
\end{equation}

\Sectionn{Appendix B}
\renewcommand{\theequation}{B.\arabic{equation}}

Here and below all vectors are taken transverse
$(D-2)$-dimensional, although  the transversality sign
$\perp$ is omitted. If the vector sign is not used, the
Minkowski metric is assumed, so that $(ab)=-\vec a \vec
b$. We use a standard representation of logarithms, i.e.
\begin{equation}
\ln a =\frac{d}{d\nu} a^{\nu}|_{\nu
=0}\label{representation of logarithm}
\end{equation}
and the Feynman parametrization
\begin{equation} \prod_{i=1}^n
a_i^{-\alpha_i} =
\frac{\displaystyle{\Gamma\left(\sum_{i=
1}^{n}\alpha_i\right)}}{\displaystyle{\prod_{i=1}^{n}\Gamma(\alpha_i)}}
\left(\prod_{i=1}^n\int_0^1dx_ix_i^{\alpha_i-1}\right)\frac{
\displaystyle{\delta(1-\sum_{i=1}^{n}x_i)}}{\displaystyle{\left(
\sum_{i=1}^{n}
a_ix_i\right)^{\displaystyle{\sum_{i=1}^{n}\alpha_i}}}}~.
\label{feynman parametrization}
\end{equation}
Using the notations
\[
R=(k_2-yk_x)^2-y\left(a_x-yb_x\right)~,\;\;k_x=k(1-x)-q_2x=q_1(1-x)-q_2
=k-q_1x~,\;\;
\]
\begin{equation}
a_x=-\left(k^2(1-x)+q_2^2x\right)~,
\;\;b_x=-k_x^2=a_x+q_1^2x(1-x)~,\;\; {\cal
J}_1=\frac{q^{\prime 2}_1}{2}J_1+(q_i\leftrightarrow
q_i^{\prime})~,
\end{equation}
from Eq.~(\ref{J1 over x}) we obtain
\[
J_1=\frac{\partial}{\partial
\nu}\int\frac{d^{2+2\epsilon}k_1}{\pi^{1+\epsilon}\Gamma(1-\epsilon)}
\frac{((q_{1}-k_1)^2)^{\nu}}{(k^2_1)^{\nu}k^2_2}\left(
\frac{q_{1}^2q_2^2}{k^2_1(q_{1}-k_1)^2}-\frac{q_2^2}{(q_{1}-k_1)^2}
-\frac{q_2^2+2k_1q_2}{k^2_1}\right)|_{\nu=0}
\]
\[
=\frac{\partial}{\partial \nu}\int_0^1
dy\int_0^1\frac{dx(1-x)^{\nu}x^{-\nu}}{\Gamma(1+\nu)\Gamma(1-\nu)}
\int\frac{d^{2+2\epsilon}k_1}{\pi^{1+\epsilon}\Gamma(1-\epsilon)}
\left(\frac{2yq_{1}^2q_2^2}{R^3}-\frac{q_2^2\nu}{(1-x)R^2}
+\frac{(q_{1}^2-k^2-2k_2q_2)\nu}{xR^2}\right)|_{\nu=0}
\]
\[
=\frac{\partial}{\partial
\nu}\int_0^1\frac{dx(1-x)^{\nu}x^{-\nu}}{\Gamma(1+\nu)\Gamma(1-\nu)}
\int_0^1
\frac{dyy^{\epsilon-1}}{\left(a_x-yb_x\right)^{1-\epsilon}}
\]
\[
\times
\left[\frac{-(1-\epsilon)q_{1}^2q_2^2}{\left(a_x-yb_x\right)}
-\frac{\nu q_2^2}{(1-x)} +\frac{\nu}{x}
(q_1^2-k^2-2yk_xq_2)\right]|_{\nu=0}
\]
\[
=\int_0^1dx \int_0^1 dyy^{\epsilon-1}
\]
\begin{equation}
\times
\left[\frac{(1-\epsilon)q_{1}^2q_2^2}{\left(a_x-yb_x\right)^{2
-\epsilon}}\ln\left(\frac{x}{1-x}\right)-\ln(1-x)\frac{d}{dx}
\frac{q_2^2}{\left(a_x-yb_x\right)^{1-\epsilon}} -\ln
x\frac{d}{dx}\frac
{q_1^2-k^2-2yk_xq_2}{\left(a_x-yb_x\right)^{1-\epsilon}}\right]~.
\end{equation}
Subsequent integrations can be performed only in the
limit $\epsilon \rightarrow 0$.  Note that proceeding to
the limit $\epsilon \rightarrow 0$ in the integrand leads
to a wrong, although convergent, integral. The
singularity at $y=0$ requires an accurate consideration.
After integration over $y$ we obtain
\[
J_1=
\int_0^1dx\left[\frac{q_{1}^2q_2^2}{a_x^2}\ln\left(\frac{x}{1-x}\right)\left(\ln
\left(\frac{a_x}{a_x-b_x}\right)+\frac{b_x}{a_x-b_x}
\right)\right.
\]
\[
\left.-\ln(1-x)\frac{d}{dx}\frac{q_2^2}{a_x}\ln
\left(\frac{a_x}{a_x-b_x}\right)-\ln x\frac{d}{dx}
\left(\frac{q_{1}^2-k^2}{a_x}-\frac{2k_xq_2}{b_x}
\right)\ln \left(\frac{a_x}{a_x-b_x}\right)\right.
\]
\begin{equation}
\left.
+\frac{1}{a_x^2}\left(q_{1}^2q_2^2\ln\left(\frac{x}{1-x}
\right)+q_2^2(k^2-q_2^2)\ln(1-x)+(q_1^2-k^2)(k^2-q_2^2)\ln
x\right)\ln a_x\right]~.
\end{equation}
Last terms come from the singularity at $y=0$.
Performing appropriate integrations by parts,
after simple though slightly tedious steps we arrive at
\[
J_1= \int_0^1dx\left(
\frac{1}{1-x}\left(\frac{q_{2}^2-q_{1}^2}{k^2}+1
\right)\ln\left(\frac{a_x}{-q_2^2}\right)\right.
\]
\begin{equation}
\left.+\frac{1}{k^2b_x}\left(q_1^2(q_{1}^2-q_{2}^2-k^2)x+k^2(q_{1}^2+q_{2}^2)-(q_{1}^2+q_{2}^2)^2
\right)\ln\left(\frac{a_x}{a_x-b_x}\right) \right) ~.
\end{equation}
Using the equalities
\[
\int_0^1\frac{xdx}{( k(1-x)- q_2 x)^2}\ln \left( \frac{
k^{ \:2}(1-x)+ q_2^{ \:2}x}{
q_1^{\:2}x(1-x)}\right)=\frac{ q_1^{\:2}+ k^{\:2}-
q_2^{\:2}}{2 q_1^{\:2}}\int_0^1\frac{dx}{( k(1-x)- q_2
x)^2}
\]
\begin{equation}
\times \ln \left(\frac { k^{ \:2}(1-x)+ q_2^{ \:2}x} {
q_1^{\:2}x(1-x)}\right) -\frac{1}{2 q_1^{\:
2}}\left(L(1-\frac{ k^{\:2}}{ q_2^{ \:2}}) -L(1-\frac{
q_2^{ \:2}}{ k^{\:2}})\right)-\frac{1}{4 q_1^{ \:2}} \ln
\left(\frac{k^{ \:2}}{ q_2^{ \:2}}\right) \ln
\left(\frac{ k^{ \:2} q_2^{ \:2}}{ q_1^{\:4}}\right)~,
\end{equation}
\begin{equation}
L(x)= -Li_2(x)= \int_0^x\frac{dy}{y}\ln
(1-y)~,
\;\;Li_2(1-x)+Li_2(1-\frac{1}{x})=-\frac{1}{2}\ln^2
x~,
\end{equation}
we obtain
\begin{equation}
J_1= \frac{4q_{1}^2q_{2}^2-(k^2-q_{1}^2-q_{2}^2)^2}{2k^2}
\int_0^1\frac{dx}{b_x}\ln\left(\frac{a_x}{a_x-b_x}\right)
+\frac{k^2+q_{2}^2-q_{1}^2}{2k^2}\ln\left(\frac{k^2}{q_2^2}\right)
\ln\left(\frac{q_1^2}{q_2^2}\right) ~. \label{J1 final}
\end{equation}

This result can be reached by another way, using the
analyticity properties of $J_1$.  To do that, let us
present $J_1$ at $\epsilon =0$ as an integral in the
Minkowski space:
\[
J_1=\int_0^{\infty}dz\int\frac{d^{2}k_1}{i\pi(k_2^2+i0)}
\left(\frac{1}{z-k_1^2-i0}-\frac{1}{z-(q_{1}-k_1)^2-i0}\right)
\]
\begin{equation}
\times\left(
\frac{q_{1}^2q_2^2}{(k^2_1+i0)((q_{1}-k_1)^2+i0)}-
\frac{q_2^2}{((q_{1}-k_1)^2+i0)}
-\frac{q_2^2+2k_1q_2}{(k^2_1+i0)}\right)~.\label{J1 in
minkowski}
\end{equation}
Here $k_1, \; q_1$ and $q_2$ are considered as vectors in
the two-dimensional Minkowski space, so that we have
\begin{equation}
d^2k_1=dk_1^{(0)}dk_1^{(1)},\;\;
k_1^{\:2}=(k_1^{(0)})^2-(k_1^{(1)})^2~,
\end{equation}
and so on.  Eq.~(\ref{J1 in minkowski}) determines $J_1$ as a
function of $q_1^2,\;q_2^{2}$ and $k^2$ for arbitrary
values of these variables. For $q_1^2= -\vec q_1^{\:2}
\leq 0,\;q_2^{2} = -\vec q_2^{\:2} \leq 0$ and $k^2=
-\vec k^{\:2} \leq 0$ Eq.~(\ref{J1 in minkowski}) turns into
the function given by  Eq.~(\ref{J1 over x}), that can be
easily seen by making the Wick rotation of the contour of
integration over $k_1^{(0)}$ and performing integration
over $z$.  At fixed negative $q_{1,2}^2$ Eq.~(\ref{J1 in
minkowski}) determines the real analytical function of
$k^2$ with the cut at $k^2\geq 0$. According to the
Cutkosky rules, one can find a discontinuity on the cut
rewriting Eq.~(\ref{J1 in minkowski}) as
\[
J_1=\int_0^{\infty}dz\int\frac{d^{2}k_1}{i\pi(k_2^2+i0)}
\left(\left(\frac{q_{1}^2q_2^2}{((q_{1}-k_1)^2+i0)}-q_2^2-2k_1q_2\right)
\left(\frac{1}{z}\left(\frac{1}{k^2_1+i0}-\frac{1}{k^2_1-z+i0)}
\right)\right.\right.
\]
\begin{equation}
\left.\left.+\frac{1}{k_1^2+i0}\frac{1}{(q_{1}-k_1)^2-z+i0}
\right)+\frac{q_2^2}{((q_{1}-k_1)^2+i0)}\left(\frac{1}{k_1^2-z+i0}
-\frac{1}{(q_{1}-k_1)^2-z+i0}\right)\right)~, \label{J1
for discontinuity}
\end{equation}
omitting the last term and making  the substitutions
(assuming $k^{(0)}\geq 0$)
\[
\frac{1}{k_2^{\:2}+i0}\frac{1}{k_1^{\:2}+i0}\rightarrow
(-2\pi i)^2
\delta(k_2^{\:2})\delta(k_1^{\:2})\theta(k_2^{(0)})\theta(k_1^{(0)})
~,\;\;\;
\]
\begin{equation}
\frac{1}{k_2^{\:2}+i0}\frac{1}{k_1^{\:2}-z+i0}\rightarrow
(-2\pi i)^2
\delta(k_2^{\:2})\delta(z-k_1^{\:2})\theta(k_2^{(0)})\theta(k_1^{(0)})~.
\end{equation}
Using these rules and removing the $\delta$-functions by
integration over $k_1$ (the most appropriate system
for this is $k^{(1)}=0,\; k^2= (k^{(0)})^2$), we obtain for the
imaginary part
\[
\Im J_1  = \pi\int_0^{\infty}dz\sum_{i=\pm}
\left[\frac{q_1^{2}q_2^{2}}{k^2}\left(
\frac{1}{\kappa^0_i(\kappa^0_i-z)}+\frac{1}{\kappa^0_iz}
-\frac{k^2\theta (k^2-z)}{z(k^2-z)\kappa_i}\right)+
\frac{k^2-q_1^{2}-q_2^{2}}{k^2}\left(
\frac{1}{\kappa^0_i-z}\right.\right.
\]
\begin{equation}
\left.\left.+\frac{1}{z}-\frac{k^2\theta
(k^2-z)}{z(k^2-z)}\right)+\frac{\kappa^0_i}{k^2(\kappa^0_i-z)}
+\frac{\kappa^0_i}{k^2z}+\frac{\kappa^0_i}{k^2(\kappa^0_i-z)}
+\frac{\theta
(k^2-z)}{(k^2-z)}\left(\frac{q_2^2}{\kappa_i}-\frac{\kappa_i}{z}\right)
\right]~,
\end{equation}
where $\kappa^0_{\pm} $ and $\kappa_{\pm} $ are given by
values of $(q_1-k_1)^2$ on the mass shells
$k_2^2=0~,\;\;k_1^2=0$ and  $k_2^2=0~,\;\;k_1^2=z$
respectively, so that
\begin{equation}
\kappa^0_{\pm}=\frac{1}{2}\left[q_1^2+q_2^2-k^2 \pm
\sqrt{(q_1^2+q_2^2-k^2)^2-4q_1^2q_2^2}\right]~, \;\;
\kappa_{\pm}=\kappa^0_{\pm}+\frac{z}{k^2}(q_2^2-\kappa^0_{\pm})~.
\end{equation}
The integration over $z$ is quite elementary and gives
\[
\Im J_1 =-\frac{\pi
}{2k^2}\left[\sqrt{(k^2-q_1^2-q_2^2)^2-4q_1^2q_2^2}
\ln\left(\frac{k^2-q_1^2-q_2^2+
\sqrt{(k^2-q_1^2-q_2^2)^2-4q_1^2q_2^2}}{k^2-q_1^2-q_2^2-
\sqrt{(k^2-q_1^2-q_2^2)^2-4q_1^2q_2^2}} \right)\right.
\]
\begin{equation}\label{Im J1}
\left.+\left(k^2-q_1^2+q_2^2\right)\ln\frac{q_1^2}{q_2^2}\right]~.
\end{equation}
The use of  this equation and  the equality (see
Refs.~\cite{gluon octet kernel} and \cite{FP2002} )
\[
\pi\sqrt{(k^2-q_1^2-q_2^2)^2-4q_1^2q_2^2}
\ln\left(\frac{k^2-q_1^2-q_2^2+
\sqrt{(k^2-q_1^2-q_2^2)^2-4q_1^2q_2^2}}{k^2-q_1^2-q_2^2-
\sqrt{(k^2-q_1^2-q_2^2)^2-4q_1^2q_2^2}} \right)
\]
\begin{equation}
=\Im\left([4q_1^2q_2^2-(k^2-q_1^2-q_2^2)^2]
\int_0^1\frac{dx}{( k(1-x)- q_2 x)^2}\ln \left( \frac{
k^{ \:2}(1-x)+ q_2^{ \:2}x}{
q_1^{\:2}x(1-x)}\right)\right)
\end{equation}
gives the result (\ref{J1 final}). The absence of a
polynomial in $k^2$ can be easily checked by considering
the integral (\ref{J1 over x}) at large $\vec
k^{\:2}\gg\vec q_1^{\:2}$. Integration regions which
could contribute in this case are $|\vec k_1|\sim |\vec
k-\vec k_1| \sim |\vec k|$ and $|\vec k_1|\sim |\vec q_1
-\vec k_1|\sim |\vec q_1|$. But the first region gives a
vanishing contribution because of the smallness of
$\ln((\vec q_1 -\vec k_1)^2/\vec k_1^{\:2})$ there.  In
the second region the integrand of Eq.~(\ref{J1 final})
with the required accuracy is anti-symmetric with respect
to the exchange $k_1\leftrightarrow q_1-k_1$, so that its
contribution vanishes as well.

\Sectionn{Appendix C}
\renewcommand{\theequation}{C.\arabic{equation}}

We start from the integrals
\[
\int_0^1dx\left(\frac{({\tilde t_1\tilde
t_1^{\prime}})^{-1}}{x(1-x)}\right)_+=\frac{L_+}{2cc^{\prime}}-\frac{c+c^{\prime}}{2cc^{\prime}}
\frac{L_-}{b-b^{\prime}}~,\;\;\;\int_0^1{dx}\left(\frac{x_2({\tilde
t_1\tilde
t_1^{\prime}})^{-1}}{x(1-x)}\right)_+=\frac{1}{2bb^{\prime}}\left[{L_+}-
\frac{b+b^{\prime}}{b-b^{\prime}} {L_-}\right]~,
\]
\[
\int_0^1{dx}\left(\frac{x_2(x_1{\tilde t_1\tilde
t_1^{\prime}})^{-1}}{x(1-x)}\right)_+=\frac{L_-}{a(b-b^{\prime})}~,\;\;\;
\int_0^1{dx}\left(\frac{x_2^2(x_1^2{\tilde t_1\tilde
t_1^{\prime}})^{-1}}{x(1-x)}\right)_+=\frac{1}{2a^2}\left[-{L_+}-
\frac{c+c^{\prime}}{b-b^{\prime}} {L_-}\right]~,
\]
\begin{equation}\label{basic integrals over x}
\int_0^1{dx}\left(\frac{x_2^2({\tilde t_1\tilde
t_1^{\prime}})^{-1}}{x(1-x)}\right)_+=\frac{1}{bb^{\prime}}\left[-1+\frac{L_+}{2bb^{\prime}}
(bb^{\prime}+a(b+b^{\prime}))
-\frac{L_-}{2(b-b^{\prime})}(c+c^{\prime}+\frac{a(b-b^{\prime})^2}
{bb^{\prime}})\right]~,
\end{equation}
where
\[
a=k_1^2~,\;\; c=(q_1-k_1)^2~,\;\;
c^{\prime}=(q^{\prime}_1-k_1)^2~,\;\;b=c-a,\;\;\;b^{\prime}
=c^{\prime}-a~,
\]
\begin{equation}\label{definition a b c L}
L_+=\ln\left(\frac{cc^{\prime}}{a^2}\right)~,\,\,
L_-=\ln\left(\frac{c}{c^{\prime}}\right)~,
\end{equation}
and after some algebra we arrive at Eq.~(\ref{J2
integrated over x}). We remind that in the integrands we
omit the terms giving zero after the subsequent
integration. Then we use the following equalities:
\[
\int\frac{d^{2+2\epsilon}k_1}
{\pi^{1+\epsilon}\Gamma(1-\epsilon)}\frac{a}{b}\ln\left(\frac{c}{a}\right)
=-\left(-q_1^2\right)^{\epsilon+1}\frac{\Gamma(1+\epsilon)
\Gamma(2+\epsilon)}{\epsilon\Gamma(4+2\epsilon)}~,
\]
\[
\int\frac{d^{2+2\epsilon}k_1}
{\pi^{1+\epsilon}\Gamma(1-\epsilon)}\frac{1}{b}\ln\left(\frac{c}{a}\right)
=\left(-q_1^2\right)^{\epsilon}\frac{\Gamma^2(1+\epsilon)}
{\epsilon\Gamma(2+2\epsilon)}~,
\]
\[
\int\frac{d^{2+2\epsilon}k_1}
{\pi^{1+\epsilon}\Gamma(1-\epsilon)}\frac{(q_1k_1)}{ab}\ln\left(\frac{c}{a}\right)
=\left(-q_1^2\right)^{\epsilon}\frac{\Gamma^2(1+\epsilon)}
{\epsilon\Gamma(1+2\epsilon)}(\psi(1+\epsilon)-\psi(1+2\epsilon))~,
\]
\[
q_1^2\int\frac{d^{2+2\epsilon}k_1}
{\pi^{1+\epsilon}\Gamma(1-\epsilon)}\frac{(q_1-k_1)
(q_1^{\prime}-k_1)}{ac}\ln\left(\frac{c}{a}\right)
=-(q_1q_1^{\prime})\left(-q_1^2\right)^{\epsilon}\frac{\Gamma^2(1+\epsilon)}
{\epsilon^2\Gamma(1+2\epsilon)}~,
\]
\[
q^2\int\frac{d^{2+2\epsilon}k_1}{\pi^{1+\epsilon}\Gamma(1-\epsilon)}
\frac{L_-}{b-b^{\prime}}
=-\left(-q^2\right)^{\epsilon+1}\frac{\Gamma^2(1+\epsilon)}
{\epsilon\Gamma(2+2\epsilon)}~,
\]
\[
\int\frac{d^{2+2\epsilon}k_1}{\pi^{1+\epsilon}\Gamma(1-\epsilon)}
\frac{c+c^{\prime}}{b-b^{\prime}}L_-
=-\left(-q^2\right)^{\epsilon+1}\frac{\Gamma^2(1+\epsilon)}
{\epsilon\Gamma(2+2\epsilon)\Gamma(3+2\epsilon)}~,
\]
\begin{equation}\label{basic integrals over k}
\int\frac{d^{2+2\epsilon}k_1}{\pi^{1+\epsilon}\Gamma(1-\epsilon)}
\frac{(q_1-k_1)
(q_1^{\prime}-k_1)}{c}\frac{L_-}{b-b^{\prime}}
=-\left(-q^2\right)^{\epsilon}\frac{\Gamma^2(1+\epsilon)}
{\epsilon\Gamma(1+2\epsilon)}(\psi(1+\epsilon)-\psi(2+2\epsilon))~.
\end{equation}
The first three of these integrals can be easily calculated with
the help of the representation
\begin{equation}
\frac{1}{b}\ln\left(\frac{a+b}{a}\right)=\int_0^1\frac{dx}{a+bx}~,
\end{equation}
the forth with the help of Eq.~(\ref{representation of
logarithm}) and the last three using the representation
\begin{equation}
\frac{L_-}{b-b^{\prime}}=\int_0^1\frac{dx}{cx+c^{\prime}(1-x)}~.
\end{equation}
Using these integrals we arrive at Eq.~(\ref{J2 at
arbitrary d}). As for the integral
$I_+(q_1,q_1^{\prime})$ of Eq.~(\ref{I_+}), it can be
written as
\[
I_+(q_1,q_1^{\prime})=-\int\frac{d^{2+2\epsilon}k_1}{\pi^{1+\epsilon}
\Gamma(1-\epsilon)}\left(\frac{(q_1-k_1)(q_1^{\prime}-k_1)}{cc^{\prime}}
\ln\left(\frac{cc^{\prime}}{(q^2)^2}\right)
-\frac{c+c^{\prime}}{cc^{\prime}}
\ln\left(\frac{k_1^2}{q^2}\right) \right)
\]
\begin{equation}
-\int\frac{d^{2+2\epsilon}k_1}{\pi^{1+\epsilon}
\Gamma(1-\epsilon)}\frac{q^2}{(q_1-k_1)^2
(q_1^{\prime}-k_1)^2}
\ln\left(\frac{k_1^2}{q^2}\right)~.\label{I+
decomposition}
\end{equation}
The first integral in Eq.~(\ref{I+ decomposition})
can be easily calculated at arbitrary$\epsilon$; we find
\[
-\int\frac{d^{2+2\epsilon}k_1}{\pi^{1+\epsilon}
\Gamma(1-\epsilon)}\left(\frac{(q_1-k_1)(q_1^{\prime}-k_1)}{cc^{\prime}}
\ln\left(\frac{cc^{\prime}}{(q^2)^2}\right)
-\frac{c+c^{\prime}}{cc^{\prime}}
\ln\left(\frac{k_1^2}{q^2}\right) \right)
\]
\[
= \frac{\Gamma^2(1+\epsilon)}
{\epsilon\Gamma(1+2\epsilon)}\left(2\left(-q^2\right)^{\epsilon}
\left(\frac{1}{\epsilon}-\psi(1)+\psi(1-\epsilon)-\psi(1+\epsilon)+
\psi(1+2\epsilon)\right)\right.
\]
\begin{equation}
\left. -\frac{1}{\epsilon}\left(\left(-q_1^2\right)^{\epsilon}+\left(-q_1^{\prime
2}\right)^{\epsilon}\right)\right)~.
\end{equation}
The second integral in Eq.~(\ref{I+ decomposition}) was
analyzed in Ref.~\cite{FFKPg}. In the limit
$\epsilon\rightarrow 0$ we have
\begin{equation}
\int\frac{d^{2+2\epsilon}k_1}{\pi^{1+\epsilon}
\Gamma(1-\epsilon)}\frac{q^2}{(q_1-k_1)^2
(q_1^{\prime}-k_1)^2} \ln\left(\frac{k_1^2}{q^2}\right)=
\frac{1}{\epsilon}\left(-q^2\right)^{\epsilon}\ln\left(\frac{(q^2)^2}
{q_1^2q_1^{\prime
2}}\right)-\frac{1}{2}\ln^2\left(\frac{q_1^2}
{q_1^{\prime 2}}\right)~.
\end{equation}

\Sectionn{Appendix D}
\renewcommand{\theequation}{D.\arabic{equation}}

We use Eqs.~(\ref{integrals Ji}), (\ref{A12}) and after a simple algebra
we obtain
\[
{\cal J}_3= \int_{0}^{1}dx
\left(\frac{1}{x(1-x)}\int\frac{d^{2+2\epsilon}k_1}
{\pi^{1+\epsilon}\Gamma(1-\epsilon)}\frac{1}{\tilde
t_1\tilde t^{\prime}_2}\left[\frac{ (D-2)}{4}
x_1x_2(q^2_{1}-2q_{1}k_{1})(q^{\prime
2}_{1}-2q^{\prime~}_{1} k_{2}) \right.\right.
\]
\[
\left.
-\frac{2x_2}{x_1}(q_1q^{\prime}_{1})(k_1(q^{\prime}_1-k_{2}))
+q^{\prime
2}_{1}\left(2\frac{q_1k_2}{x_1}-2q_1q^{\prime}_{1}+q_1k_{1}\right)
+4x_1q_1^2(q^{\prime}_{1}k_2)+q^{\prime}_{1}q_{1}(q^{\prime}_{1}q_{1}-
q^{\prime}_{1}k_{1}-q_{1}k_{2}) \right.
\]
\[
\left.
+2(q^{\prime}_{1}k_{1})(q_{1}k_2)-2(q^{\prime}_{1}k_{2})(q_{1}k_1)
+\frac{q^{\prime
2}_{1}}{k^2_{2}}\left(2\frac{x_2}{x_1}(q_1k_2)(q^{\prime}_{1}k_{1}
-kk_2)\right.\right.
\]
\begin{equation}
\left.\left.\left.
+x_2(q^{\prime}_{1}k_{2})(q_2^2-k^2+2q_1k_2)+2(q_2k_2)
(q_1q+q_1k_2)\right)+q^{\prime
2}_{1}q^{2}_{1}\frac{(q_2k_2)(q^{\prime}_{2}k_{1})}{k^2_{1}
k^2_{2}}\right]\right)_+ +(q_i\leftrightarrow
q^{\prime}_i)~. \label{J3 begin}
\end{equation}
Terms in Eq.~(\ref{J3 begin}) with $x_1$ in the denominators
require subtraction, so that the prescription
(\ref{integral +}) is important for them (and only for
them). Note that neither the integrand in Eq.~(\ref{J3 begin})
itself, nor the subtraction terms contain non-integrable
infrared singularities. Nevertheless, the integral ${\cal
J}_3$ has a pole at $\epsilon =0$. The pole comes from the
ultraviolet region. Evidently, the ultraviolet divergency
is artificial and appears as a result of the separation
shown in Eq.~(\ref{As=sumAi}). It is easy to see from
Eqs.~(\ref{A11}) and (\ref{A12})
that the terms leading to such divergencies cancel
in the sum $A_2+A_3$.

The integration  over $k_1$ is performed using a standard
Feynman parameterization. The basic integrals are
\[
\int\frac{d^{2+2\epsilon}k_1}
{\pi^{1+\epsilon}\Gamma(1-\epsilon)}\frac{1}{\tilde
t_1\tilde
t^{\prime}_2}=x_1x_2\int_0^1\frac{dz}{Q^{2(1-\epsilon)}}~,\;\;\;
\int\frac{d^{2+2\epsilon}k_1}
{\pi^{1+\epsilon}\Gamma(1-\epsilon)}\frac{k^{\mu}_i}{\tilde
t_1\tilde t^{\prime}_2}=x_1x_2\int_0^1\frac{dz
r^{\mu}_i}{Q^{2(1-\epsilon)}}~,\;\;\;
\]
\begin{equation}\label{integrals with t1t2}
\int\frac{d^{2+2\epsilon}k_1}
{\pi^{1+\epsilon}\Gamma(1-\epsilon)}\frac{k^{\mu}_1k^{\nu}_2}{\tilde
t_1\tilde
t^{\prime}_2}=x_1x_2\int_0^1\frac{dz}{Q^{2(1-\epsilon)}}
\left(r^{\mu}_1r^{\nu}_2-\frac{g^{\mu\nu}}
{2\epsilon}Q^2\right)~,
\end{equation}
where
\[
Q^2=-x(1-x)(q_1^2 z+q_1^{\prime 2}(1-z))-z(1-z)(q_2^2
x+q_2^{\prime 2}(1-x)-q^2x(1-x))~,
\]
\begin{equation}\label{definition Q2,p1,p2}
r_1=zxq_1+(1-z)(xk-(1-x)q_2^{\prime})~,\;\;r_2=z((1-x)k-xq_2)
+(1-z)(1-x)q_1^{\prime}; \;\; r_1+r_2=k~.
\end{equation}
The integrals with $\tilde t_1\tilde t^{\prime}_2k_i^2$
can be calculated joining first $\tilde t_1$ and $\tilde
t^{\prime}_2$ and then $k_i^2$. We have
\[
\int\frac{d^{2+2\epsilon}k_1}
{\pi^{1+\epsilon}\Gamma(1-\epsilon)}\frac{1}{\tilde
t_1\tilde
t^{\prime}_2k_i^2}=-x_1x_2(1-\epsilon)\int_0^1dz\int_0^1
\frac{ydy}{\left(y\left(\mu_i^2+yr_i^2\right)\right)
^{2-\epsilon}}~,
\]
\[
\int\frac{d^{2+2\epsilon}k_1}
{\pi^{1+\epsilon}\Gamma(1-\epsilon)}\frac{k^{\mu}_i}{\tilde
t_1\tilde
t^{\prime}_2k_i^2}=-x_1x_2(1-\epsilon)\int_0^1dz\int_0^1
\frac{dy
y^2r^{\mu}_i}{\left(y\left(\mu_i^2+yr_i^2\right)\right)
^{2-\epsilon}}
\]
\[
=-x_1x_2\int_0^1\frac{dz r^{\mu}_i}{\mu_i^2 Q^2}+{\cal
O}(\epsilon)~,
\]
\[
\int\frac{d^{2+2\epsilon}k_1}
{\pi^{1+\epsilon}\Gamma(1-\epsilon)}\frac{k^{\mu}_ik^{\nu}_i}{\tilde
t_1\tilde
t^{\prime}_2k_i^2}=-x_1x_2(1-\epsilon)\int_0^1dz\int_0^1
\frac{ydy }{\left(y\left(\mu_i^2+yr_i^2\right)\right)
^{2-\epsilon}}\left(y^2r^{\mu}_ir^{\nu}_i-
\frac{g^{\mu\nu}y\left(\mu_i^2+yr_i^2\right)}
{2(1-\epsilon)}\right)
\]
\begin{equation}\label{integrals with t1t2k2}
=-x_1x_2\int_0^1dz\left[\frac{r^{\mu}_ir^{\nu}_i}{r_i^2}\left(\frac{1}
{r_i^2}\ln\left(\frac{Q^2}{\mu_i^2}\right)-\frac{1}{Q^2}\right)
-\frac{g^{\mu\nu}}{2r_i^2}\ln\left(\frac{Q^2}{\mu_i^2}\right)\right]
+{\cal O}(\epsilon)~,
\end{equation}
where
\[
\mu_i^2=Q^2-r_i^2~,\;\;\;\mu_1^2=-zxq_1^2-(1-z)(xk^2+(1-x)q_2^{\prime
2})~,\;\;
\]
\begin{equation}
\mu_2^2=-z((1-x)k^2+xq_2^{2})-(1-z)(1-x)q_1^{\prime 2}~.
\end{equation}
Finally, the integral with $\tilde t_1\tilde
t^{\prime}_2k_1^2k_2^2$ can be calculated joining $\tilde
t_1$ and $\tilde t^{\prime}_2$, then the result with
$k_1^2$ and subsequently with $k_2^2$. We obtain
\[
\int\frac{d^{2+2\epsilon}k_1}
{\pi^{1+\epsilon}\Gamma(1-\epsilon)}\frac{k_2^{\mu}k_1^{\nu}}{\tilde
t_1\tilde
t^{\prime}_2k_1^2k_2^2}=x_1x_2(2-\epsilon)(1-\epsilon)
\int_0^1\int_0^1\int_0^1\frac{dzydyt^3dt}{\left(t\left(y\left(
\mu_1^2+2kr_1\right)-k^2+t(k-yr_1)^2\right)\right)^{3-\epsilon}}
\]
\[
\times \left[(k-yr_1)^\mu (k-t(k-yr_1))^\nu
+\frac{g^{\mu\nu}}{2(2-\epsilon)}\left(y\left(
\mu_1^2+2kr_1\right)-k^2+t(k-yr_1)^2\right)\right]
\]
\[
=x_1x_2\int_0^1\frac{dz}{d}\left[-k^{\mu}k^{\nu}\left(\frac{1}{k^2}
+\frac{Q^2}{d}{\cal
L}\right)-r_2^{\mu}k^{\nu}\left(\frac{1}{\mu_2^2}
-\frac{\mu_1^2}{d}{\cal
L}\right)-k^{\mu}r_1^{\nu}\left(\frac{1}{\mu_1^2}
-\frac{\mu_2^2}{d}{\cal L}\right)\right.
\]
\begin{equation}\label{integrals with t1t2k1k2}
\left.+r_2^{\mu}r_1^{\nu}\left(\frac{1}{Q^2}
+\frac{k^2}{d}{\cal L}\right)+\frac{g^{\mu\nu}}{2}{\cal
L}\right] +{\cal O}(\epsilon),
\end{equation}
where
\[
d=\mu_1^2\mu_2^2+k^2Q^2=z(1-z)x(1-x)\left((k^2-q_1^2-q_2^{\prime
2})(k^2-q_1^{\prime
2}-q_2^2)+k^2q^2\right)+q_1^2q_2^2xz(x+z-1)
\]
\begin{equation}\label{definition d and L in appendix}
+q_1^{\prime 2}q_2^{\prime 2}(1-x)(1-z)(1-x-z),\;\;\;
{\cal L}=\ln\left(\frac{\mu_1^2\mu_2^2}{-k^2Q^2}\right).
\end{equation}
At arbitrary $D$ we have
\[
{\cal J}_3= \int_{0}^{1}dx\int_0^1
dz\left\{-\frac{1+\epsilon}{\epsilon}q_1q_1^{\prime}\left(x_1x_2Q^{2\epsilon}
+\frac{2}{x_1}\left(x_2Q^{2\epsilon}-Q_0^{2\epsilon}\right)\right)\right.
\]
\[
\left.+\frac{1+\epsilon}{2Q^{2(1-\epsilon)}}
x_1x_2(q^2_{1}-2q_{1}r_{1})(q^{\prime
2}_{1}-2q^{\prime~}_{1}
r_{2})-\frac{2}{x_1}\left[\left(x_2q_1q^{\prime~}_{1}(r_1(q^{\prime}_{1}
-r_2))-q^{\prime
2}_{1}q_1r_2\right)\frac{1}{Q^{2(1-\epsilon)}}\right.\right.
\]
\[
\left. + \left(z(1-z)q^{\prime 2}_{2}q_1q^{\prime~}_{1}
+q^{\prime
2}_{1}(zq_1k+(1-z)q_1q^{\prime}_{1})\right)\frac{1}{Q_0^{2(1-\epsilon)}}
\right]+\frac{1}{Q^{2(1-\epsilon)}}\left(q^{\prime
2}_{1}q_1\left(r_{1}-2q^{\prime}_{1}\right)\right.
\]
\[
\left.\left. +4x_1q_1^2(q^{\prime}_{1}r_2)+q^{\prime}_{1}q_{1}(q^{\prime}_{1}
q_{1}-q^{\prime}_{1}r_{1}-q_{1}r_{2})+2(q^{\prime}_{1}r_{1})
(q_{1}r_2)-2(q^{\prime}_{1}r_{2})(q_{1}r_1)\right)
\right.
\]
\[
\left. +{q^{\prime
2}_{1}}\int_0^1ydy\left[\frac{1}{\left(y(\mu_2^2+yr_2^2)\right)^{1
-\epsilon}}
\left(-\frac{x_2}{x_1}q_1(q^{\prime}_{1}+k)+x_2q_1q^{\prime}_{1}
+q_1q_2\right)\right.\right. -\frac{y(1-\epsilon)}{\left(y(\mu_2^2+yr_2^2)\right)^{2
-\epsilon}}
\]
\[
\times\left(2\frac{x_2}{x_1}(q_1r_2)(q^{\prime}_{1}k-y(q^{\prime}_{1}+
k)r_2)+x_2(q^{\prime}_{1}r_2)(q_2^2-k^2+2yq_1r_2)
+2(q_2r_2)(q_1(q+yr_2))\right)
\]
\[
+\frac{1}{\left(y(\mu_0^2+yr_0^2)\right)^{1 -\epsilon}}
\left(\frac{1}{x_1}q_1(q^{\prime}_{1}+k)\right)
+\frac{y(1-\epsilon)}{\left(y(\mu_0^2+yr_0^2)\right)^{2
-\epsilon}}
\left(\frac{2}{x_1}(q_1r_0)(q^{\prime}_{1}k-y(q^{\prime}_{1}+
k)r_0)\right)
\]
\[
+(2-\epsilon)(1-\epsilon)
q^{2}_{1}\int_0^1\frac{t^3dt}{\left(t\left(y\left(
\mu_1^2+2kr_1\right)-k^2+t(k-yr_1)^2\right)\right)^{3-\epsilon}}
\left[(q_2(k-yr_1))
(q_2^{\prime}(k-t(k-yr_1)))\right.
\]
\begin{equation}
\left.\left.\left.
+\frac{q_2q_2^{\prime}}{2(2-\epsilon)}\left(y\left(
\mu_1^2+2kr_1\right)-k^2+t(k-yr_1)^2\right)\right]\right]\right\}
+(q_i\leftrightarrow q^{\prime}_i)~, \label{J3 at
arbitrary d}
\end{equation}
where
\begin{equation}
Q_0^2=-z(1-z)q^{\prime 2}_2,\;\;\;
\mu_0^2=-zk^2-(1-z)q^{\prime
2}_1,\;\;\;r_0=zk+(1-z)q_1^{\prime},\;\;\;r^2_0=Q_0^2-\mu_0^2.
\label{definition mu0 Q0 p02}
\end{equation}
In the limit $\epsilon \rightarrow 0$ we arrive at
Eq.~(\ref{J3 at d=4}).

\end{document}